\documentclass[aps,twocolumn,amsmath,amssymb,notitlepage,superscriptaddress,longbibliography,nofootinbib]{revtex4-1}

\usepackage[T1]{fontenc}

\usepackage{amsmath}
\usepackage{amssymb}
\usepackage{bm}
\usepackage{graphicx}
\usepackage{xcolor}
\definecolor{darkblue}{rgb}{0,0,0.6}
\definecolor{darkred}{rgb}{0.6,0,0}
\usepackage[%
colorlinks=true,
urlcolor=darkblue,
linkcolor=darkred,
citecolor=darkblue
]{hyperref}

\usepackage{wasysym}
\usepackage{lmodern}
\usepackage{colonequals}
\usepackage{dcolumn}

\renewcommand\vec{\mathbf}

\newcommand{\sss}{\vec{s}}

\newcommand{\mcE}{\mathcal{E}}
\newcommand{\mcV}{\mathcal{V}}
\newcommand{\mcF}{\mathcal{F}}

\graphicspath{{./Figures/}}

\begin{document}
	
	\title{
		Active hydraulics laws from frustration principles
	}
	
	\author{Camille Jorge}
	\affiliation{Univ. Lyon, ENS de Lyon, Univ. Claude Bernard, CNRS, Laboratoire de Physique, F-69342, Lyon.}
	\author{Am\'elie Chardac}
	\affiliation{Univ. Lyon, ENS de Lyon, Univ. Claude Bernard, CNRS, Laboratoire de Physique, F-69342, Lyon.}
	\affiliation{Department of Physics, Brandeis University, Waltham, MA, 02453, USA}
	\author{Alexis Poncet}
	\affiliation{Univ. Lyon, ENS de Lyon, Univ. Claude Bernard, CNRS, Laboratoire de Physique, F-69342, Lyon.}
	\author{Denis Bartolo}
	\affiliation{Univ. Lyon, ENS de Lyon, Univ. Claude Bernard, CNRS, Laboratoire de Physique, F-69342, Lyon.}
	\email{denis.bartolo@ens-lyon.fr}

	\begin{abstract}
		Viscous flows are laminar and deterministic. Robust linear laws accurately predict their streamlines in structures as complex as blood vessels, porous media and 
		pipes networks. 
		However, biological and synthetic active fluids defy these fundamental laws. 
		Irrespective of their microscopic origin, confined active flows are intrinsically bistable, and therefore non-linear.  
		As a consequence, their emergent patterns in channel networks are out of reach of available theories, and lack quantitative experiments.
		Here,  we lay out the basic laws of active hydraulics. 
		We  show that active hydraulic flows  are non-deterministic and yield degenerate streamline patterns ruled by frustration at nodes with an odd coordination number.
		More precisely,  colloidal-roller experiments in trivalent networks reveal how active-hydraulic flows realize dynamical spin ices. 
		The resulting streamline patterns split into two distinct classes of self-similar loops, which  reflect the fractionalization of topological defects at the subchannel scales.
		Informed by our measurements, we   formulate the laws of active hydraulics as a double spin model.  
		A series of mappings on loop O($n$) models then allow us to  exactly predict the  geometry of  the degenerate streamlines.
		We expect our fundamental understanding to provide robust  design rules for active microfluidic devices, and to offer unanticipated avenues to understand the motion of living cells and organisms in complex habitats.
	\end{abstract}

\let\oldaddcontentsline\addcontentsline
\renewcommand{\addcontentsline}[3]{}
	
	\maketitle
	
	When they invented irrigation, the ancient civilizations of Egypt and Mesopotamia relied on the first fundamental rule of hydraulics: mass conservation~\cite{Rost}. 
	In our modern language we formulate it as the first Kirchhoff's law.
	In steady state, the sum of the fluxes  vanishes at
	each node of a channel network ($\sum_j \Phi_{ij}=0$, where $\Phi_{ij}$ is the flux from node $j$ to node $i$). 
	It is only eight millennia later, that  Hagen, Poiseuille and Darcy discovered the second law of hydraulics~\cite{Hagen1839,Poiseuille1844,Darcy1856}.
	Darcy's law relates the mass fluxes to pressure gradients:  $\Phi_{ij}=-K_{ij}(P_i-P_j)$, where $P_i$ is the fluid pressure and $K_{ij}$ the hydraulic conductance.
	Given these two linear laws, and a set of boundary conditions, we can  predict the viscous flows of any hydraulic network, regardless of its geometrical complexity.
	Although it can be computationally challenging to solve and optimize, viscous hydraulics is deterministic and fully predictable.
	
	However, over the past decade physicists, chemists and material scientists have complexified this neat picture. They engineered active fluids that escapes the fundamental laws of hydraulics~\cite{Marchetti2013}.
	Either by hacking biological engines, such as molecular motors and bacteria, or by motorizing synthetic soft materials, we have learned how to power fluids at the scale of their elementary building blocks. 
	When confined, the resulting active materials enjoy spontaneous laminar flows even in the absence of any external drive or boundary motion~\cite{Bricard2013,WiolandNJP2016,Zhang2017,Wu2017,Doostmohammadi2018,Saintillan2018rheology,Opathalage2019}.  
	In channels and pipes, active fluids flow in one direction or the other with the same probability, and can even resist opposing pressure gradients~\cite{Morin2018}. 
	In other words, they are intrinsically bistable~\cite{Woodhouse2017,Wu2017,Morin2018}.
	This fundamental deviation from Darcy's law makes active hydraulics a challenging problem whose basic laws are yet to be determined. 
	
	From an experimental perspective, active hydraulics has been mostly restricted to  straight channel geometries and closed loops~\cite{WiolandNJP2016,Wu2017,Hardouin2022}. 
	To the best of our knowledge, the investigation of interconnected channel networks has remained limited to pioneering experimental demonstrations in simple one-node networks~\cite{Wu2017}. 
	More significant progress have been made on the theory front, in particular by Dunkel, Woodhouse and coworkers, who realized  the computational power of Active Fluidic Networks (AFN) and their dynamical multistability~\cite{Woodhouse2016stochastic,Woodhouse2017,Forrow2017,Woodhouse2018information}.
	
	In this article, we perform large-scale active hydraulics experiments. 
	We show that spontaneous laminar flows are frustrated in networks including nodes with an odd coordination number.
	Focusing on fully frustrated networks, we show that the resulting active flows realize dynamical spin ices with extensively degenerate flow patterns.
	We show that unlike passive fluids, the random streamline geometry depends on the aspect ratio of the elementary channels.
	We explain this polymorphism by combining experiments and numerical simulations, and show that it originates from topological-defect fractionalization at the subchannel level. 
	We then elucidate the self-similar geometry of the flow patterns by mapping them on the frustrated structures of magnetic spin ices, and of the so-called loop O($n$) models~\cite{Moessner2006,Nisoli2013,OrtizAmbriz2019,Udagawa2021}.
	Altogether our findings allow us to identify the full set of laws ruling the steady flows of active matter circulating through interconnected channels networks.
	
	\begin{figure*}
		\includegraphics[width=\textwidth]{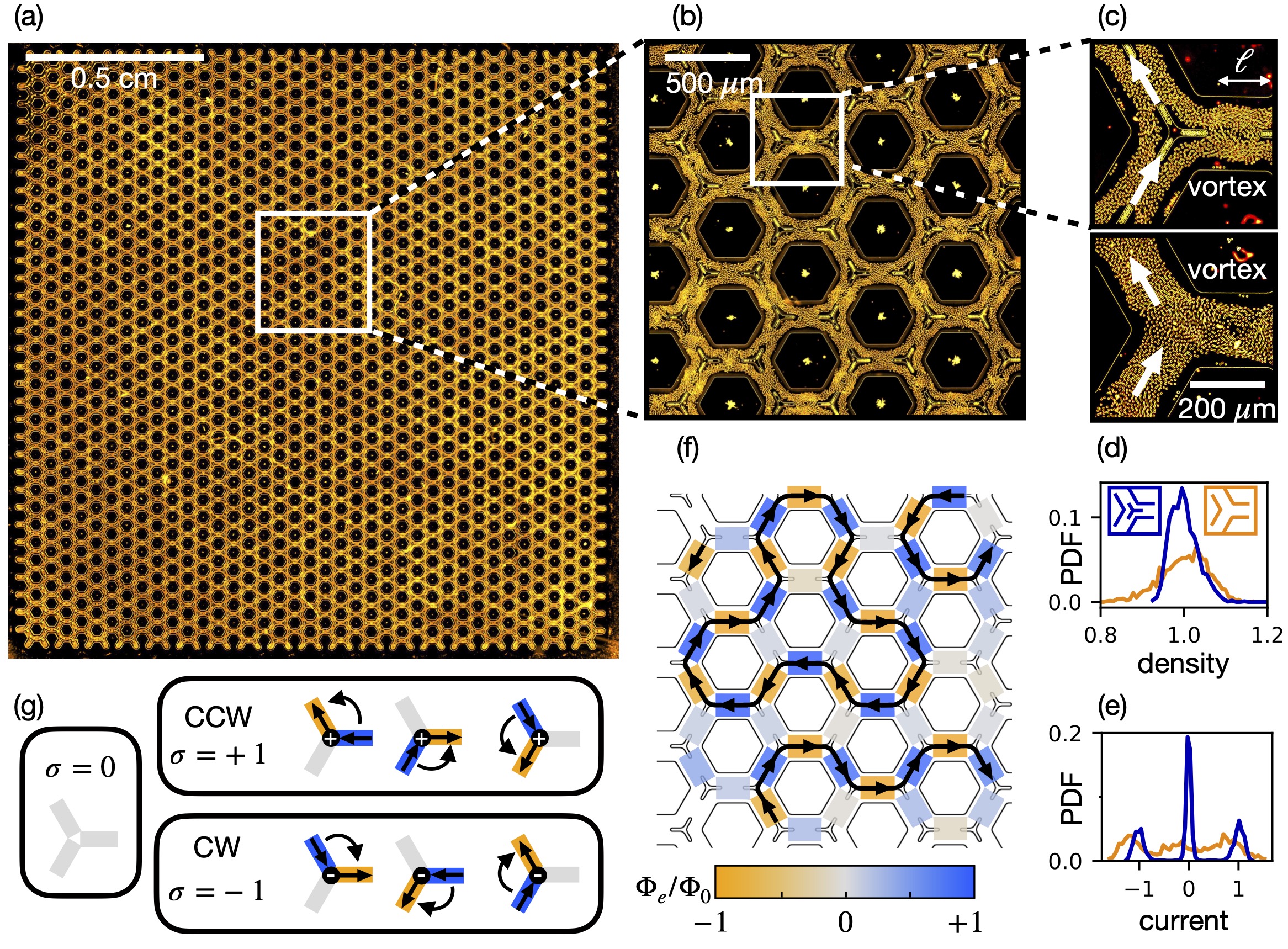}
		\caption{{\bf Active hydraulics in trivalent networks}
			{\bf (a)} Macroscopic picture of a trivalent microfluidic network. It encloses a two-dimensional active fluid made of flocking Quincke rollers. 
			{\bf (b)} Close-up view. We can distinguish steady vortices in a finite fraction of the channels.
			{\bf (c)} All the results reported in this article do not depend on the specific geometry of the nodes that connect the microfluidic channels. 
			Top: node including a star-shape flow splitter. 
			Bottom: Node with no splitter. 
			In steady state only two of the three channels support laminar flows. The third (horizontal) channel hosts a steady vortex. 
			{\bf (d)} Distribution of the roller density in the channels. The density is homogeneous whatever the node geometry but features less fluctuations when including a star-shape splitter.
			{\bf (e)} Distribution of the current $\Phi_e$ normalized by the most probable value $\Phi_0=400 \, \rm s^{-1}$. We find a trimodal distribution for both node geometries.
			{\bf (f)} The heatmap shows the local value of the current $\Phi_e$ supported by each edge of the channel network.  
			The sign of $\Phi_e$ is arbitrarily defined using the bipartite nature of the honeycomb lattice.
			$\Phi_e$ is chosen to be positive when the flow goes from a given sublattice to the other.
			The solid line indicate the shale of the streamlines and the arrows the orientation of the flow. 
			The data correspond to the same region of space showed in {\bf (b)}.
			{\bf(g)} Classification of the seven possible vertices at each node. 
			The color of the edges indicate the sign of the current. It is associated to a classical spin-1 variable  $\Phi_e\in{-1,\,0,\,1}$, ($\Phi_e=0$ means: no current on the edge $e$). 
			We distinguish three different classes of vertices indexed by another spin variable $\sigma\in{-1,\,0,\,1}$ that distinguishes the  handedness of the three-color patterns. $\sigma$ is defined at the nodes of the network.
			$\sigma_i=+1$  when the set of colors at the node $i$ is a positive permutation of $\{\textrm{orange, blue, gray} \}$ indexed in the clockwise direction. $\sigma_i=-1$ for negative permutations, and $\sigma_i=0$ when all three fluxes vanish, see also Methods.  
		}
		\label{Fig1}
	\end{figure*}
	\begin{figure*}
		\centering
		\includegraphics[width=\textwidth]{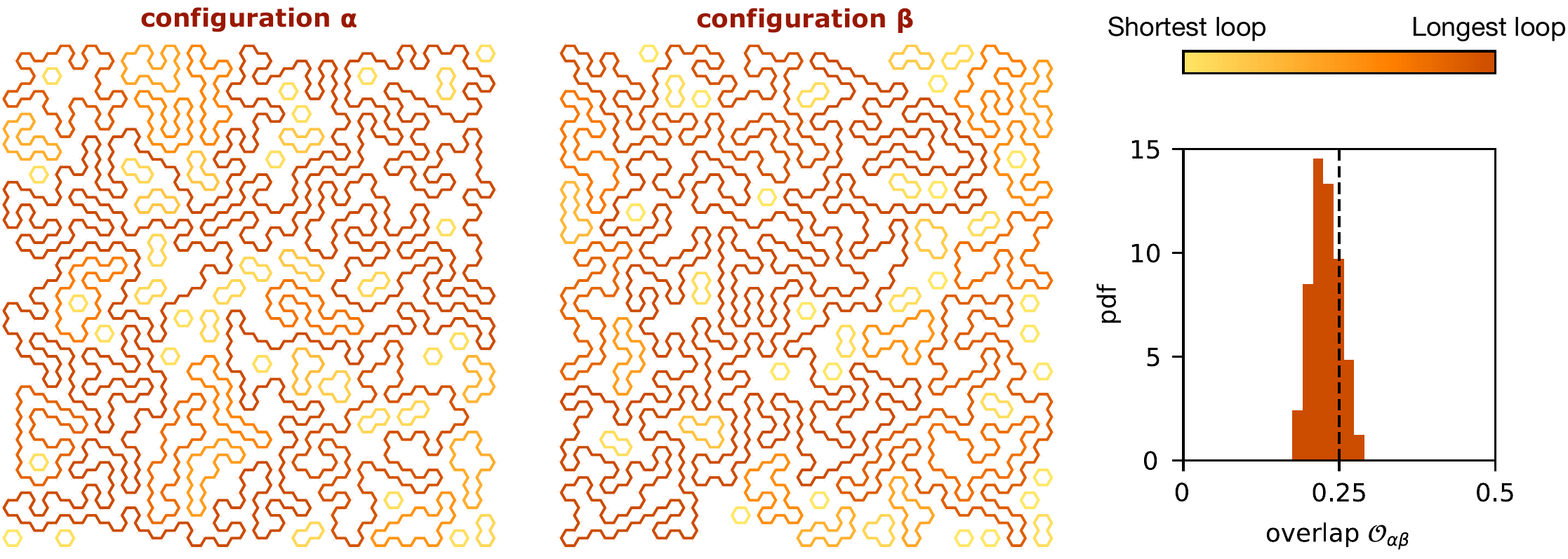} 
		\caption{
			{\bf Degeneracy of the streamline patterns.} Two realizations of the same experiment  ($\epsilon=0.7$) lead to two different streamline patterns with markedly different geometries. See Fig.S7 in SI for more realizations.
			In all realizations the streamlines form a soup of self-avoiding loops colored according to their length (the darker the longer).
			We stress that the overlap distribution computed over ten independent experiments indicates very little correlations from one realization to another. The overlap function $\mathcal{O}_{\alpha \beta}$ quantifies the difference between the flow patterns  in the two realizations $\alpha$ and $\beta$ of the same experiment:   $\mathcal{O}_{\alpha \beta} = \langle 1 - |\mathbf \Phi_\alpha(e) - \mathbf \Phi_\beta(e)|\rangle_e$, where $\Phi(e)=\pm 1,0$ is the sign of the flux in the edge $e$ of the honeycomb graph, see Methods. 
			The typical value of the overlap is 0.2, which is close to the value expected for uncorrelated random current fields. 
			In the limit case where the $\Phi_\alpha(e)$ are uncorrelated, we can perform a crude approximation. Ignoring spatial correlations, we define the probability $1-2p$ that $\Phi(e)=0$ and assume that $\Phi(e)=\pm 1$ have the same probability $p$. 
			The average overlap between two random configurations would then be $\langle\mathcal{O}_e^{\alpha \beta}\rangle = 1-4p (1-p)$. Estimating $p$ from the fraction of edges supporting no net current, we find $p\sim0.25 - 0.3$, which yields $\langle\mathcal{O}_e^{\alpha \beta}\rangle = 0.25 - 0.16$.
			%
		}
		\label{Fig2}
	\end{figure*}
	
	To tackle the problem of active hydraulics experimentally, we need a model active fluid, and model hydraulic networks, see Fig.~\ref{Fig1}. 
	We provide all the technical details about our experiments in the Method section. 
	In short, our active fluid is a flocking liquid assembled from colloidal Quincke  rollers~\cite{Bricard2013}, which features bistable laminar flows in straight channels~\cite{Morin2018}. 
	Here, we confine the roller fluid in networks of channels. They form a  honeycomb structure, which includes up to 3,200 trivalent nodes as illustrated in Fig.~\ref{Fig1}a. 
	In their pioneering experiments Wioland \textit{et al.} showed that active flows can strongly depend on the specific  geometry of their boundaries~\cite{Wioland2016}.
	To establish the robustness of our findings, we therefore consider two different node geometries where the trivalent junctions either include a star-shape splitter or not, see Figs.~\ref{Fig1}c. 
	We also systematically vary the aspect ratio $\epsilon=w/\ell$ of the roller steams  in the  channels, by  
	varying their length $\ell$  from $120~\mu \rm m$ to $200~\mu \rm m$, see Fig~\ref{Fig1}c. 
	The channel width $w$ is kept constant and equal to $200\,\rm \mu m$, about a hundred times larger than the colloids' radius ($2.4\,\rm \mu m$).
	
	We first observe a transient dynamics where the active fluid undergoes strong density and velocity fluctuations  over system spanning scales, see Supplementary Video~1.
	It then reaches a  steady state where the density is homogeneous across the whole device, see Fig.~\ref{Fig1}a and \ref{Fig1}d, and Supplementary Videos 2 and 3.
	By contrast, the mass-current distribution in the channels is heterogeneous, Fig.~\ref{Fig1}e. 
	It is trimodal with two symmetric peaks associated to spontaneous flows at constant speed, and a higher peak revealing the presence of channels supporting no net current, see Fig.~\ref{Fig1}e and~\ref{Fig1}f.
	We find that the peaks of the distributions of the density and current ($\Phi_e$) are narrower when using the splitter geometry shown in Fig.~\ref{Fig1}c. 
	All the results reported below therefore correspond to this node geometry. 

	Unlike in disconnected straight channels, the active flows are never laminar throughout the whole sample: at least $40\%$ of the channels host steady vortices and therefore support no net mass flux, see Fig.~\ref{Fig1}b, \ref{Fig1}e and \ref{Fig1}f.
	This series of observations stems from the geometrical frustration of active laminar flows. 
	It can be understood as follows. 
	As in most active fluids, the spontaneous flows of the Quincke rollers  operate at constant speed. 
	When the fluid density is homogeneous, the magnitude of the net fluxes $|\Phi_{e}|$ measured along all the network edges ($e$) is peaked on a constant value $\Phi_0$ (in our experiments $\Phi_0= 400\,\rm s^{-1}$), Fig. \ref{Fig1}e. 
	The active flows are therefore frustrated at any node with an odd coordination.
	Flowing at constant speed and conserving mass are two incompatible constraints that cannot be simultaneously met in steady state.
	In the specific case of trivalent network, frustration is accommodated at each node by the emergence of vortices in either one, or three, edges where the flux vanishes ($\Phi_{e}=0$), see Supplementary videos 4, 5 and 6.
	Irrespective of the  channel aspect ratio, the situation where all three fluxes vanish is also possible but less likely to happen as seen in Supplementary Figure 1.

	The geometry of the streamlines is therefore determined by the conflicting imperatives set by activity and mass conservation. 
	The resulting local frustration defines a set of seven possible flow rules at the vertices which we classify in  Fig.~\ref{Fig1}g. 
	From a condensed matter perspective, they are akin to the spin-ice rules responsible for the ground state degeneracy of  magnetic textures in frustrated magnets~\cite{Moessner2006,Nisoli2013,OrtizAmbriz2019,Udagawa2021}. 
	More specifically, the six most probable vertices of Fig.~\ref{Fig1}g define a three-coloring model on the honeycomb lattice~\cite{Baxter1970,Baxter2016}. 
	This first analogy with spin-ice physics  explains the vast degeneracy of the flow patterns found in our experiments. 
	Active hydraulic flows are not deterministic. Repeating the same experiment in the same periodic geometry, we observe a plethora of disordered flow patterns. We illustrate them in Fig.~\ref{Fig2} and show that they hardly feature any correlation.

	\begin{figure*}
		\centering
		\includegraphics[width=0.95\textwidth]{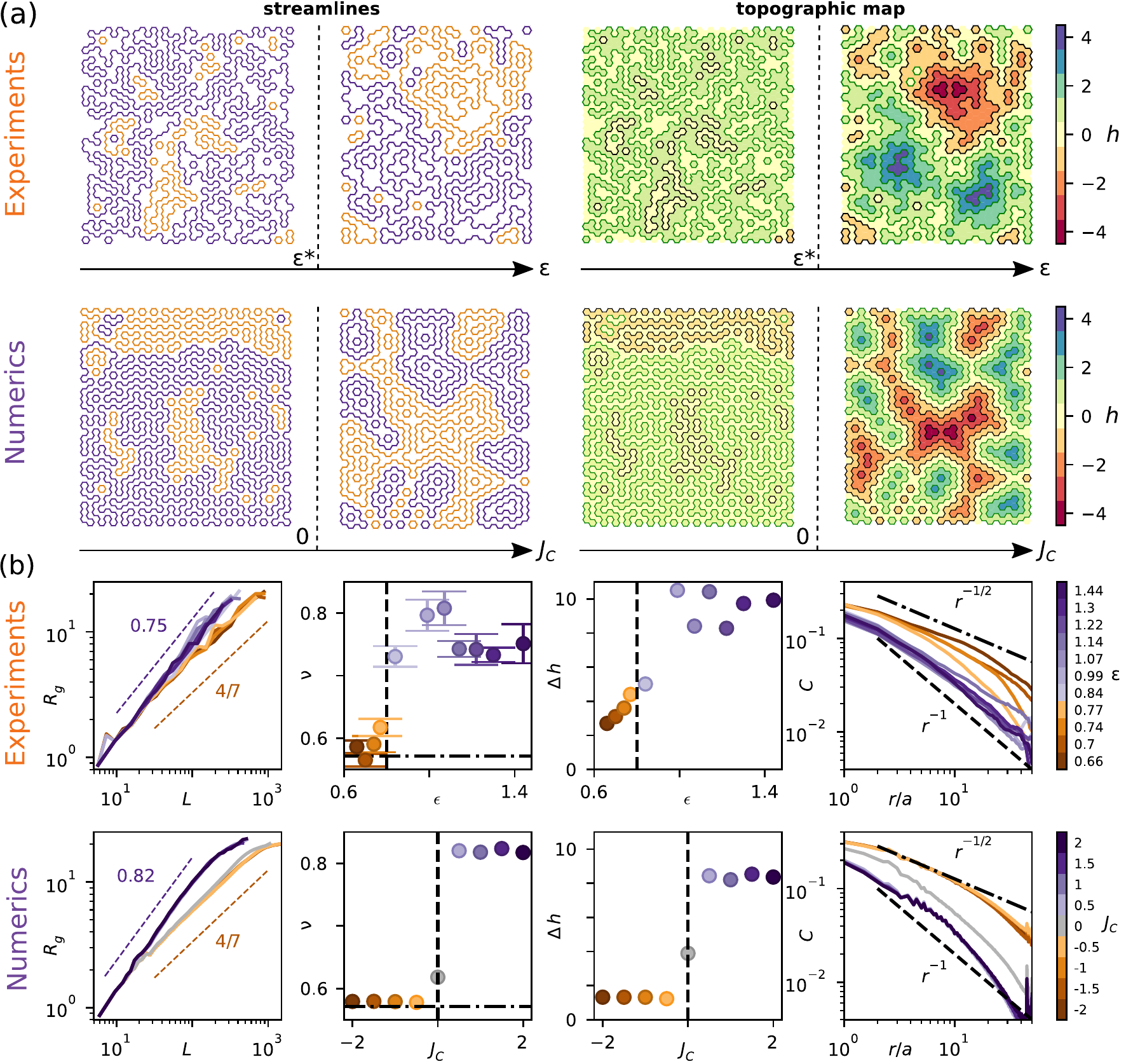}
		\caption{
			{\bf Polymorphism of the streamlines.}
			{\bf (a)} Evolution of the streamline morphology with the channel aspect ratio. The color indicates the flow direction along the loops (Orange: clockwise, purple: counterclockwise).
			Top row: Experiments.
			The two set of streamline loops correspond to $\epsilon=0.7$ and $\epsilon=1.1$ respectively. 
			Below $\epsilon^\star=0.8$, the streamlines are crumpled and segregated. 
			By contrast, above $\epsilon^\star$, they form nested and persistent loop patterns. 
			The two heat maps are the topographic map of the same experiments. The local height field is defined as the sum of the winding numbers associated to all loops winding around a given point in space, see Methods.  Below $\epsilon^\star$ the segregation of the streamlines translate in a two-level topography. Conversely, above $\epsilon^\star$ the  map features high hills and deep valleys.
			Bottom row: Numerical simulations.
			Streamlines and topographic maps predicted by the low energy configurations of the Hamiltonian defined by Eq.~\eqref{eq:H} for $J_C=-2$ and $J_C=2$. The qualitative agreement between the simulations and experiments confirm the relevance of our active-hydraulics laws. It also shows that $J_C$ is a good proxy for the channel aspect ratio.
			{\bf (b)} Characterization of the streamline geometries: experiments, simulations and theory.
			All simulations are performed on a honeycomb graph having the same size and boundary conditions as in the experiments.
			First column: For all values of $\epsilon$ (experiments), and $J_C$ (simulations), the gyration radius $R_g$ grows algebraically with the length of the streamline loops $N$: $R_{\rm g}\sim N^\nu$.
			The dashed lines are the two theoretical predictions detailed in SI based on loop O($n$), and three-coloring models.
			Second column: Variations of $\nu$ with $\epsilon$ (experiments) and $J_C$ (simulations). Dashed-dotted line: exact  prediction in the small $J_C$ and large $J_A$ limit (closely packed O(1) loop model).
			Dashed line: exact  prediction  for $J_C=0$ and  $J_A\gg1$ limit (closely packed O(2) loop model).
			Third column: Nesting level quantified by the  maximal height difference on the topographic maps (average over ten realizations). The nesting level jumps at $\epsilon=\epsilon^\star$ (experiments), and $J_C=0$ (simulations)
			Fourth column: $C(r)$ is the probability to find two nodes at a distance $r$ within the same loop. When $\epsilon>\epsilon^\star$ the correlation $C(r)$ decays much faster than when $\epsilon<\epsilon^\star$. The trend is even clearer in the simulations, where the curves collapse on  exponential and  power law mastercurves. Dash-dotted line: Exact  prediction in the limit $J_C\ll0$, dashed line: exact  prediction for $J_C=0$.
		}
		\label{Fig3}
	\end{figure*}
	The streamlines form nearly close-packed ensembles of self-avoiding loops. 
	This geometry is a direct consequence of the spin-ice rules sketched in Fig~\ref{Fig1}g, where the seven vertices are the generators of the self-avoiding random walks on the honeycomb lattice. 
	The closing of the streamlines into loops readily follows from mass conservation. 
	We note that these patterns are consistent with the AFN model numerically discussed in~\cite{Woodhouse2016stochastic,Woodhouse2018information} for degree-3-vertex graphs.  
	However the loops' morphology  strongly depends on the aspect ratio of the channels  $\epsilon$, Figs.~\ref{Fig3}a.
	Obviously, this polymorphism cannot be explained by the sole spin-ice rule of Fig.~\ref{Fig1}g which is agnostic to the channel lengths.
	To gain more insight on the origin of streamline patterns, we  quantitatively characterize the loops' geometry.
	Whatever the channel aspect ratios, the loops are self-similar: their gyration radius $R_{\rm g}$ grows algebraically with their length $L$, Fig.~\ref{Fig3}b.
	For small $\epsilon$ values, the loops are collapsed and segregated, by contrast, when $\epsilon$ is large, the streamlines are more persistent and form nested structures.
	We quantify these observations by plotting the  exponent $\nu$ of the gyration radius and see that it undergoes a sharp increase when $\epsilon$ exceeds $\epsilon^\star=0.8$, Fig.~\ref{Fig3}c. 
	
	The transition from segregated to nested loops occurs at the same value $\epsilon^\star$. 
	To see this, we classically  identify the orientated streamlines with the contour lines of the height fields $h$  of rough landscapes~\cite{Nienhuis1982}. 
	We show in Fig.~\ref{Fig3}a the corresponding topographic maps of $h$, see also Methods. 
	We then quantify the  level of nesting by the maximal height difference $\Delta h$ which increases sharply at $\epsilon^\star$,
	Fig.~\ref{Fig3}b.

	\begin{figure*}
		\includegraphics[width=\textwidth]{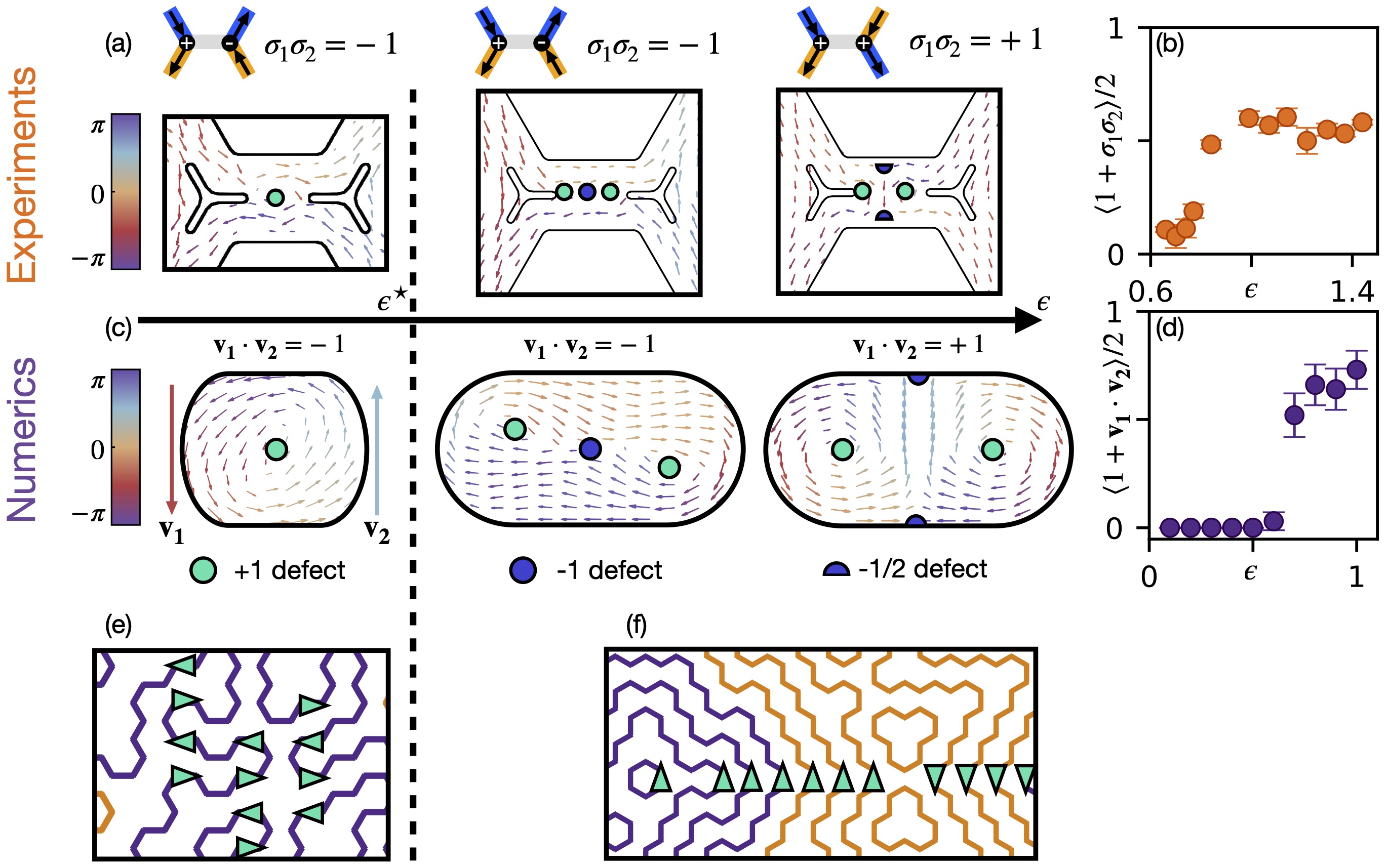}
		\caption{{\bf Orientational interactions between adjacent streamlines and topological defect fractionalization.} 
			{\bf (a)} Closeup on the flow field in two adjacent streamlines separated by a  channel that supports  no net flux. 
			Below $\epsilon^\star$ the central channel hosts a single vortex. 
			Above $\epsilon^\star$  the central channel hosts two vortices. They can either rotate in the same or opposite directions.
			When they rotate in the same direction, the flow field includes a $-1$ topological charge separating the two $+1$ charges at the center of the vortices. Conversely, when the two vortices rotate in opposite directions the $-1$ charge is fractionalized into two $-1/2$ defects bound to the channel walls.
			{\bf (b)} Fraction of parallel contacts across each channel supporting no net flow plotted against the aspect ratio $\epsilon$. $\epsilon$ is defined in Fig. 1 for the experiments, and is given by the ratio between the length and the width of the anisotropic simulation boxes.
			The fraction of parallel contacts is equal to $\frac 1 2 (1+\sigma_1\sigma_2)$, where $\sigma_1\sigma_2=1$ when the two vertical flows are along the same direction and $-1$ otherwise.  
			The flow couplings are mostly antiparallel below $\epsilon^\star$ and mostly parallel above $\epsilon^\star$. 
			{\bf (c)} Numerical resolution of the Toner-Tu equations in cigar-shaped boxes, see SI for details about the numerical method and material parameters. We denote $w$ the width of the box, $\ell$ the length of the straight line joining the half-circles, the aspect ratio of the boxes is defined by $\epsilon = \ell / w$.
			In agreement with what we observe in {\bf b}, increasing the aspect ratio of the anisotropic box result in a transition form a flow field with a single $+1$ topological charge to flow fields including two vortices, which can be either separated by a single $+1$ defect or two fractionalized $-1/2$ defects bound to the walls to conserve the overall topological charge.  
			{\bf (d)} Repeating the same simulation with random initial conditions we find that the  steady states where the rightmost ($v_2$) and leftmost ($v_1$) flows are parallel prevail in boxes with large aspect ratios. 
			In other words the transition from antiparallel to parallel contacts in our experiments are not specific to colloidal rollers but generic to flocking matter (i.e Toner-Tu fluids).
			{\bf (e)}
			Close-up on the experimental streamlines measured in a channel where $\epsilon<\epsilon^\star$. Close-up on the experimental streamlines measured in a channel where $\epsilon>\epsilon^\star$. Orange :  streamline loops flowing in the clockwise direction, purple : streamline loops flowing in the counterclockwise direction. 
			The green arrow indicates the local orientation of the flows.
			The streamlines are crumpled and segregated. The antiparallel couplings between adjacent streamlines promote this geometry. 
			{\bf (f)}
			Same as in {\bf (e)} for $\epsilon>\epsilon^\star$.  
			The streamlines are persistent and nested. The parallel couplings between adjacent streamlines promote this geometry. 
		}
		\label{Fig4}
	\end{figure*}
	These observations prompt us to investigate deeper the orientational interactions between the streamlines.
	To do so, we measure the fraction of parallel configurations when two adjacent streamlines are separated by a channel supporting no net flux, Fig.~\ref{Fig4}b.
	To measure this quantity we note that it is given by  $\langle 1 + \sigma_1 \sigma_2 \rangle /2$ where the spin variables $\sigma_i$  measure the handedness of the vertices  
	see Fig.~\ref{Fig1}g and Fig.~\ref{Fig4}a.

	Fig.~\ref{Fig4}b 
	shows that antiparallel contacts ($\sigma_1 \sigma_2 =-1$) prevail when $\epsilon<\epsilon^\star$, whereas most of the contacts are parallel ($\sigma_1\sigma_2 = +1$) when $\epsilon>\epsilon^\star$.
	This central result indicates that active hydraulic flows are not only shaped by the spin-ice rules but also by short-range orientational interactions between adjacent streamlines.
	To elucidate their nature, we investigate the morphology of the vortical flows at the subchannel scale, Fig.~\ref{Fig4}a.
	We find a clear structural change at $\epsilon^\star$. 
	Below $\epsilon^\star$, the channels with no net flux ($\Phi_{e}=0$) host a single vortex. 
	This vortex couples the adjacent channels as a gear would couple two circulators.
	The continuity of the flow field then favors antiparallel couplings between neighboring streamlines.
	This short range coupling is consistent with collapsed streamline loops including a number of hairpins, see   Fig.~\ref{Fig4}e.

	The situation where $\epsilon>\epsilon^\star$ is more subtle. 
	We find that the zero-flux channels host two vortices, Fig.~\ref{Fig4}a. 
	In most channels, they rotate in opposite directions, Fig.~\ref{Fig4}b. The continuity of the flow field therefore promotes parallel couplings, which explains the emergence of  nested structure of the streamlines as clearly seen in Fig~\ref{Fig4}f. 
	However in Fig.~\ref{Fig4}a we can also see configurations with two co-rotating vortices which promote antiparallel couplings. 
	We now need to understand whether the dominance of counter-rotating vortices is specific to our experiments, or generic to polar active fluids.
	We address this question numerically. 
	We model the active flows using Toner-Tu  hydrodynamics of polar active matter~\cite{Toner1995,Toner1998}, and solve these generic  equations using a finite element solver, see Methods. 
	We consider the simple geometry of an anisotropic cigar-shaped chamber with tangent boundary conditions, see  Figs.~\ref{Fig4}c.
	A systematic investigation would go beyond the scope of this article, we therefore focus on a single set of material parameters consistent with earlier measurements in Quincke-roller fluids~\cite{Geyer_2019,Chardac2021}. 
	As the channel aspect ratio increases, we observe a clear transition between one-vortex and two-vortex configurations. 
	As a vortex hosts a $+1$ topological charge, when two vortices are present, an additional $-1$ topological defect must coexist with them to conserve the overall $+1$ topological charge~\cite{Chardac2021}. 
	Remarkably, we find that the most likely configuration does not correspond to the coexistence of $+1$ and $-1$ charges, Fig~\ref{Fig4}d. 
	By contrast, we observe a fractionalization of the $-1$ charge into two $-1/2$ singularities, Fig~\ref{Fig4}c.
	The fractional charges are stuck on the long channel walls and stabilize the counter-rotation of the two $+1$ vortices. 
	This numerical result is in excellent agreement with our experimental findings Fig~\ref{Fig4}a.
	We therefore conclude that the prevalence of parallel couplings between the macroscopic streamlines (Fig.~\ref{Fig4}b) originates from defect fractionalization in the flow field, and does not rely on the specifics of Quincke rollers.

	We are now equipped to state the laws of active hydraulics and map them on a double spin model:
	(i) The currents $\Phi_{e}$ on the edges $e=\{i, j\}$ can only take three values: ${-1,0,+1}$. It is a classical spin-1 variable. 
	(ii) The stability of the laminar active flows in confined geometry penalizes the $\Phi_e=0$ state. 
	(iii) In steady state, mass conservation imposes the  spin-ice constraint $\sum_{j}\Phi_{ij}=0$ which frustrates (ii) at all nodes having an odd coordination number, Fig.~\ref{Fig1}g.
	(iv) Finally, the aspect ratio of the channels  results in effective (anti)ferromagnetic interactions between adjacent streamlines. 
	
	To express these four elementary laws in more quantitative terms, we identify the streamlines with the equilibrium configurations of an effective Hamiltonian $\mathcal H$ that couples two spin variables: $\Phi_e=\pm1,\,0$   , and $\sigma_i=\pm1,\,0$ which defines the handedness of the $i^{\rm th}$ node  (see Fig.~\ref{Fig1}g and caption):
	\begin{equation}
		\mathcal H=-J_{\rm A}\sum_{\langle i,j\rangle}\Phi_{ij}^2
		-J_{\rm C}\sum_{\langle i,j\rangle}\delta_{\Phi_{ij},0}\sigma_i\sigma_j.
		\label{eq:H}
	\end{equation}
	The first term reflects the activity of the fluid. $J_A$ is a positive constant, which penalizes the states of vanishing flows (Law ii). 
	The second  term reflects the orientational interactions between adjacent streamlines  (Law iv). 
	$J_C$ is a proxy for the channel aspect ratio:
	$J_C<0$  corresponds to antiparallel couplings, \textit{i.e.} to channels where $\epsilon<\epsilon^\star$,  conversely $J_C>0$  corresponds to parallel couplings, \textit{i.e.} to channels where $\epsilon>\epsilon^\star$, see Figs. \ref{Fig4}a, \ref{Fig4}e and \ref{Fig4}f.
	We stress that $J_{C}$ only couples  vertices connected by channels that support no net flux.
	We minimize $\mathcal H$ using a Monte Carlo worm algorithm~\cite{Barkema1998} which enforces the spin-ice rule imposed by mass conservation (Law iii). 
	We then compare the equilibrium states  to our experimental measurements in Fig.~\ref{Fig3}b.
	The excellent  agreement between the computed and measured streamline geometries confirms the predictive power of our active hydraulic laws.

	We can now gain a deeper insight by exploiting quantitative analogies with a series of statistical field theories that were hitherto unrelated to any experimental system. 
	Let us first discuss he limit of strong parallel interactions ($J_C\gg0$), which is the most complex to analyze. This difficulty stems from the subexetensive degeneracy of the ground state. 
	This observation was first made on the three coloring model introduced in Ref.~\cite{Verpoort2018} which maps on Eq.~\eqref{eq:H}. 
	The ground state configurations  indeed correspond to fully nested patterns of streamlines, which we never find in our experiments and simulations. 
	In fact, we probe metastable states for which no exact prediction is available.  
	We however fully characterize them in SI, and show in Fig.~\ref{Fig3}b that our model correctly accounts for our experimental measurements in finite-size systems. 
	
	Mappings to field theories are more fruitful in the other two limits $J_{\rm C}\ll 0$ and $J_{\rm C}=0$. In these cases, our spin model allows exact predictions of the streamline geometry.
	In the limit of small aspect ratios, viz.
	$J_{\rm C}\ll 0$, we can unambiguously define a net circulation around each face of the honeycomb lattice, see SI Fig.~S9b. 
	The handedness of the circulation around each face defines yet another classical spin variable  $\Omega=\pm1$, which corresponds to $2|h|-1$, see Figs.~\ref{Fig3}a and~\ref{Fig3}b.
	Altogether, the constraints imposed by $J_A\gg0$ and $J_C\ll0$ translate into
	antiferromagnetic couplings between the adjacent $\Omega$ spins. 
	As a result, in this regime, the streamlines are nothing else but the domain walls between regions of opposite magnetization in the ground states of a frustrated Ising antiferromagnet, see Fig.~\ref{Fig3}b and SI Fig.~S9b.
	We  use this analogy to provide exact expressions for the gyration radius, the nesting level and the probability $C(r)$ to find two edges at a distance $r$ in the same loop, Fig.~\ref{Fig3}b.
	The domain walls of the Ising antiferromagnet (AF) are obviously segregated and the nesting level is given by $\Delta h=1$~\cite{Verpoort2018}. 
	The exact predictions of the gyration radius exponent and of the decay of $C(r)$ are more complex to compute. 
	They rely on the mapping of AF Ising models on the loop $O(1)$  theory~\cite{Blote1994}, which we recall in SI. 
	Using this mapping, we find  $\nu=4/7$~\cite{Kondev1995,Kondev1996} and $C(r)\sim r^{-1/2}$.
	These three exact predictions  agree with our experimental findings, and numerical simulations, in the limit $\epsilon<\epsilon^\star$ ($J_C<\ll0$), see Fig.~\ref{Fig3}b and SI.  
	Finally, when $J_C=0$, the microscopic defect fractionalization does not affect the streamline orientations, and our model reduces to Baxter's three coloring model~\cite{Baxter1970}.
	It consists in coloring the edges of a honeycomb lattice as sketched in Fig.~\ref{Fig1}g.
	Baxter's model map on an exactly solvable statistical model of interacting loops known as the O(2) loop model, which we recall in SI as well.
	The loops of this model identify to the streamlines. 
	Using this second powerful analogy we find that the  gyration radius exponent should be $\nu_0=2/3$~\cite{Kondev1996}, and that $C(r)\sim r^{-1}$.
	Remarkably these two exact predictions agree with the values measured in our experiments and simulations at the crossover between the parallel and antiparallel regimes~Fig.~\ref{Fig3}c.
	Altogether the agreement between our experiments and theories  establish the predictive power of our frustrated-spin model which accurately describes the random geometry of frustrated active-hydraulic flows. 
	
	As a final remark, we stress that the four local laws of active hydraulics apply broadly, beyond the specifics of periodic lattices and  polar active matter. They should describe the flows of any form of active fluid animated by spontaneous laminar flows in complex networks, from cell tissues, to bacteria suspensions, to active gels and liquid crystals.
	We therefore expect our findings to provide 
	a robust set of design rules for active microfluidic devices and offer new insights on the dynamics of groups of living cells and animals in heterogeneous environments  and complex habitat~\cite{Hansell2005,Perna2017,Martinez2021}.
	
	{\bf Acknowledgements.} We thank A. Morin and D. Geyer for help with preliminary experiments. We also thank D. Carpentier, P. Holdsworth and L. Jaubert for insightful comments. 
	This work was supported by ANR grant WTF and ERC SPAM.

	CJ, AC and AP have equally contributed to this work. D. B. designed the project. A. C. and C. J. performed the experiments. C. J. and A. 
	P. performed the finite element simulations. A. P. worked out the theory and performed the Monte-Carlo simulations. All authors discussed the results and wrote the manuscript.

\let\addcontentsline\oldaddcontentsline
\clearpage
\widetext

\begin{large}
	\begin{center}
		\textbf{
			Active hydraulics laws from frustration principles
		}
	
		\medskip
		\textbf{Supplementary Information}
	\end{center}
\end{large}

\setcounter{secnumdepth}{3}
\renewcommand{\theequation}{S\arabic{equation}}
\renewcommand{\thefigure}{S\arabic{figure}}
\setcounter{equation}{0}
\tableofcontents

\section{Experimental methods}

\subsection{Quincke rollers experiments.}

The experimental setup was described in~\cite{morin2017,geyer2018}. 
In short, we use polystyrene colloids of radius
$2.4~\mu \rm m$
(Thermo Scientific G0500) dispersed in a solution of hexadecane including $5.5 \times 10^{-2}$ wt \% of AOT salt (Dioctyl sulfosuccinate sodium salt).
The microfluidic devices are made of two electrodes spaced by a $25$-$\rm \mu m$-thick double-sided tape. 
We pattern the bottom surface of the device with  channel networks.
They are made of a $2$-$\rm \mu m$-thick layer of insulating photoresist resin (Microposit S1818) patterned by means of standard UV lithography as explained in~\cite{morin2017}. 
In all our experiments the networks have the shape of a honeycomb structure, see Fig.~\ref{hexa_piv}a.  
The width of the channels is $200 ~\rm \mu m$ and we vary their lengths from $120 ~\rm \mu m$ to $220 ~\rm \mu m$. 
The overall size of the channel network is in the order of $1.5 \times 1.5 \rm ~cm^2$. 

The electrodes are glass slides coated with ITO (indium tin oxide, Solems, ITOSOL30, thickness $80 ~\rm nm$). 
We start the experiments by filling  the microfluidic chambers homogeneously with the colloidal solution, we then let the colloids sediment on the bottom electrode.
The average packing fraction of the colloidal monolayer is approximately equal to $30$\% in all our experiments. 
We operate at this packing fraction to make sure that our active fluid remains  in the polar-liquid state, far from the flocking-transition
threshold (approximately equal to $0.1$\%)~\cite{Bricard2013}. 
We then apply a DC voltage of $110~\rm V$ to trigger the Quincke instability and cause the colloids to roll at a constant speed $v_0 \approx 0.8 ~\rm mm.s^{-1}$ on the bottom electrode. Each experiment is repeated from six to ten times.

\subsection{Impact of the node geometry on the emergent flow patterns}
\begin{figure}[h!]
	\includegraphics[width=\textwidth]{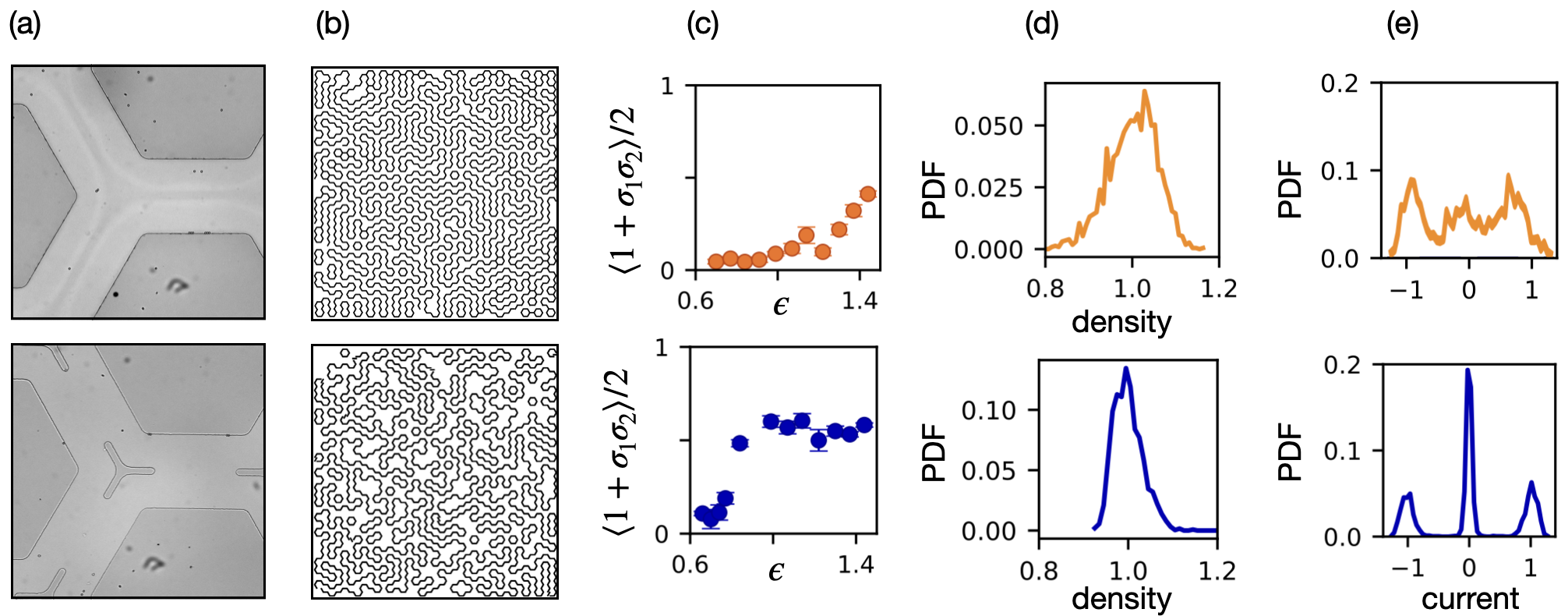}
	\caption{\textbf{(a)} Zoom on the two junctions geometries. Top: simple trivalent junction. Bottom: trivalent junction with a splitter. 
		\textbf{(b)} Top: streamlines in a networks with simple trivalent nodes. $\epsilon = 0.7$. Bottom: streamlines in a networks with splitters. $\epsilon = 0.7$. Streamlines form self avoiding loops, but splitters forbid crossings between streamlines. 
		\textbf{(c)} Fraction of parallel couplings for simple nodes (top) and for nodes with splitters (bottom). 
		\textbf{(d)} Distribution of the average density in channels. Top : simple nodes, $
		\epsilon = 0.84$. Bottom: nodes with splitters,  $
		\epsilon = 0.84$.
		\textbf{(e)} Distribution of the average current in channels.Top : simple nodes, $
		\epsilon = 0.84$. Bottom: nodes with splitters,  $
		\epsilon = 0.84$.}
	\label{hexa_piv}
\end{figure}

We use two different designs for the nodes of the network.
The first design is a simple junction between three straight channels, whereas the second design includes a flow splitter at each node, Fig.~\ref{hexa_piv}a. 
Our  results do not qualitatively depend on the specific geometry of the nodes: the streamlines form self-avoiding loops, Fig.~\ref{hexa_piv}b, and the fraction of ferromagnetic couplings increases with the aspect ratio of the channels, Fig.~\ref{hexa_piv}c. 
However, including flow-splitters results in more homogenous fluid flows, Fig.~\ref{hexa_piv}d, and reduce the crossings between the streamlines, see Figs.~\ref{hexa_piv}b. 
In the main text, all the results correspond to the splitter geometry.

\subsection{Measurement of the velocity fields}
\begin{figure}[h!]
	\includegraphics[width=\textwidth]{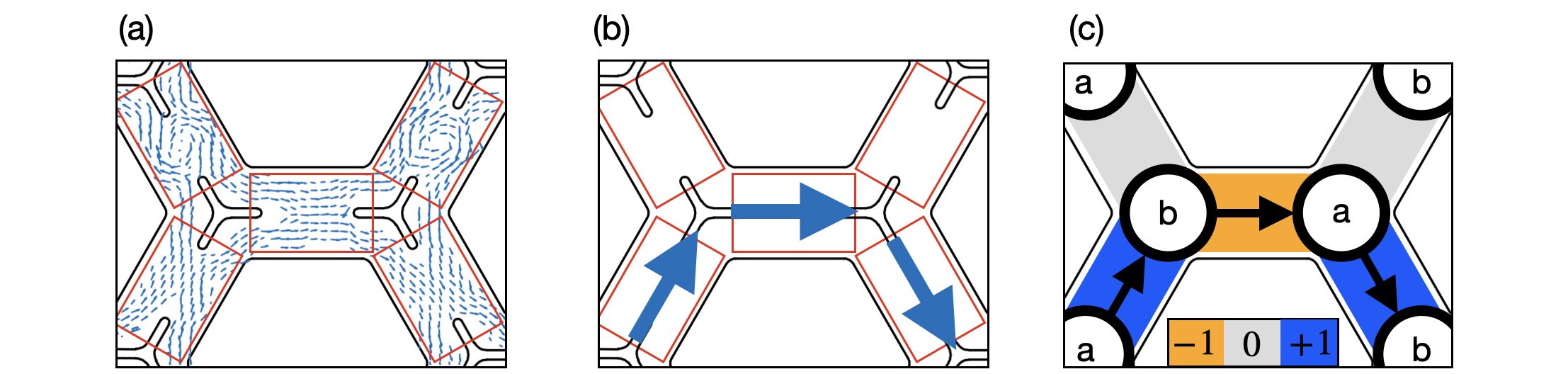}
	\caption{\textbf{(a)} Velocity field. The averaging windows are sketched in red. 
		\textbf{(b)} Sketch of the averaged velocity in each averaging window. 
		\textbf{(c)} Scalar value of the velocity. It is $+1$ (resp. $-1$) when the fluid flows on average from a `a'-node to a `b'-node (resp. from a `b'-node node to a `a'-node). }
	\label{SIFig2}
\end{figure}

Once the active flow reaches a steady state, we image the whole network for $5~\rm s$ with a Nikon AZ100 microscope.
We record the videos with a Luxima LUX160 camera (Ximea) at a frame rate of 200 fps. 
To measure the velocity field  $\mathbf{v}$ we use a conventional particle-imaging velocimetry (PIV). In practice, we use the PIVLAB MATLAB package~\cite{thielicke2014}.
The PIV box size is $48 \times 48 \rm ~ \mu m^2$, the PIV boxes overlap  over a half of their size, Fig.~\ref{SIFig2}a and \ref{SIFig2}b.
Before constructing the streamlines, we average the velocity field over the width of each channel as shown in Fig.\ref{SIFig2}. 
We define the average velocity in each channel as a scalar quantity. 
We use the bipartite geometry of the honeycomb lattice to define its sign. 
Our sign convention is more easily explained using the sketch of Fig.~\ref{SIFig2}c.
Denoting by `a' and `b' the two sublattices,
we choose to assign a $+$ sign (blue color) to the velocity when the fluid flows from a `a'-node to a `b'-node and a $-$ sign (orange color) when the fluid flows from a `b'-node to a `a'-node.

\subsection{Density and current fields.}
\begin{figure}[h!]
	\includegraphics[width=\textwidth]{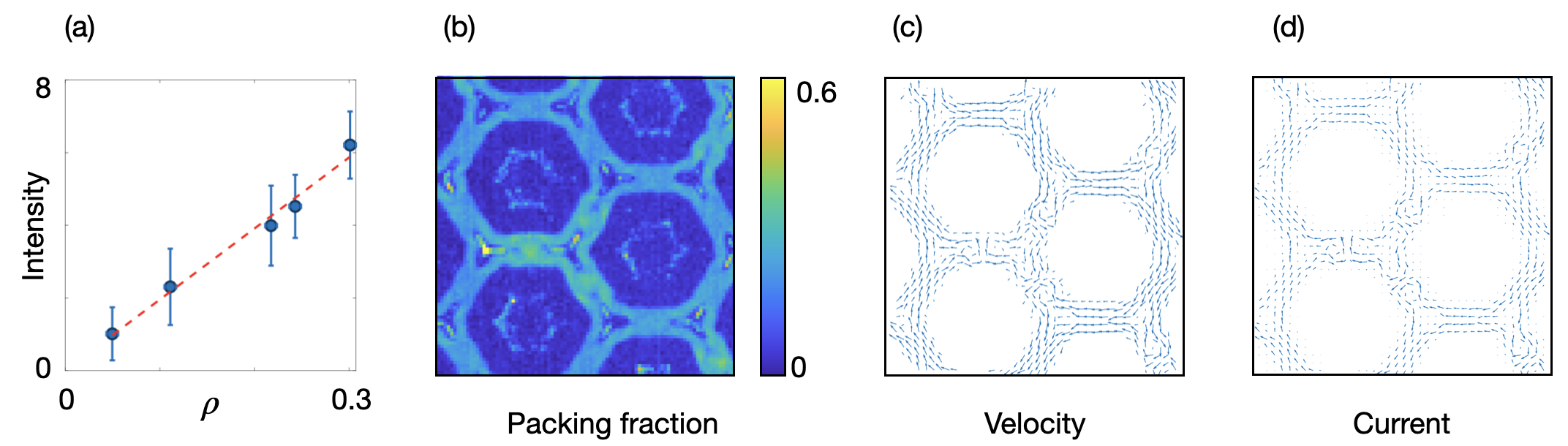}
	\caption{
		\textbf{(a)} Fluorescence intensity with respect to the packing fraction $\rho$. 
		\textbf{(b)} Colormap of the packing fraction field. 
		\textbf{(c)} Velocity field. 
		\textbf{(d)} Current field.}
	\label{SIFig3}
\end{figure}

To measure the local current fields $\mathbf{j}(\mathbf r)\equiv\rho(\mathbf r)  \mathbf{v}(\mathbf r)$, we measure the density  and the velocity field in two different movies of the same experiment. 
For both measurements, we use a Hamamatsu ORCA-Quest qCMOS camera mounted on a Nikon AZ100 microscope with a $2 \times $ objective. 
As we cannot image the whole network  at once, we  perform multiple acquisitions and stitch our images, which is always possible in the steady state.
To measure the density field, we perform epifluorescence imaging at 10 fps, and use  the local fluorescence intensity  as a proxy for the colloid density. 
Fig.~\ref{SIFig3}a shows that, as expected, both quantities are proportional 
to one another. 
The calibration was performed
on static images  using a higher magnification and using a conventional particle detection algorithm (ImageJ). 

We then use bright field imaging and record at a higher frame rate (200 fps) to perform our PIV analysis. 
The PIV box size is $24 \times 24 \rm ~ \mu m^2$. 
We then average both density, $\rho$, and velocity fields $\mathbf v$ over time and we multiply them to reconstruct the local current field $\mathbf j$ see Figs~\ref{SIFig3}b, \ref{SIFig3}c and \ref{SIFig3}d.
Finally we measure the average current $\Phi_{ij}$ along the edges of the network by averaging $\mathbf{j}$ over the channel joining the nodes $i$ and $j$.

\section{Data analysis}
In this section we detail how we analyse our data. We 
use the same methods and algorithms for our experimental and numerical data.

\subsection{Construction of the streamlines}
To construct the streamlines we first define a discrete current field $\Phi_e$ on the edge $e$ connecting the nodes $i$ and $j$.
$\Phi_e$ can take three different values: $\pm 1$ in channels that support a net current and $0$ in channel hosting vortices.
In practice we construct the $\Phi_e$ field as follows. 
We measure the average scalar product between the local velocity and the unit vector pointing in the direction of a channel. The average is performed over the channel area.  When this quantity is larger than $0.5$ we set $\Phi_e=+1$, when it is smaller than $-0.5$, we set $\Phi_e=-1$, and $\Phi_e=0$ otherwise. 
In our simulations $\Phi=\pm1,0$ by definition (see section \ref{ss:spinmodel}).

Once the $\Phi_e$ are defined we can unambiguously construct the oriented streamlines along the principal axis of the honeycomb lattice.
We then use a depth-first search algorithm to detect and label each individual loop in the streamlines soup~\cite{Cormen2022}.
%
Once the individual 
loops are identified, we can readily measure their gyration radius and the probability $C(r)$ that two edges of the lattice separated by a distance $r$ belong to the same loop (see section \ref{ss:compa}).

\subsection{Constructing a topographic map from oriented streamlines}

\label{ss:height}

We now explain how to quantify the nesting of the streamlines.
To do so, we first convert the ensemble of oriented loops into a topographic map.
We note $\mcF$ the ensemble of the hexagonal faces of the honeycomb network.
Considering a configuration $\{\Phi_e\}$ of the current field on the edges, we define a height field $h_f$ on the faces.

The loops are oriented, we can then define the winding number of each loop around a given face $f$, it is equal to $\pm 1$ when the face is lassoed by the loop and $0$ otherwise.
The height field of the topographic map is then  
defined at a face of the network as the sum of the winding numbers of all loops winding around this point. As a reference, the height is taken to be zero outside the network. 

\begin{figure}
	\centering
	\includegraphics[width=\textwidth]{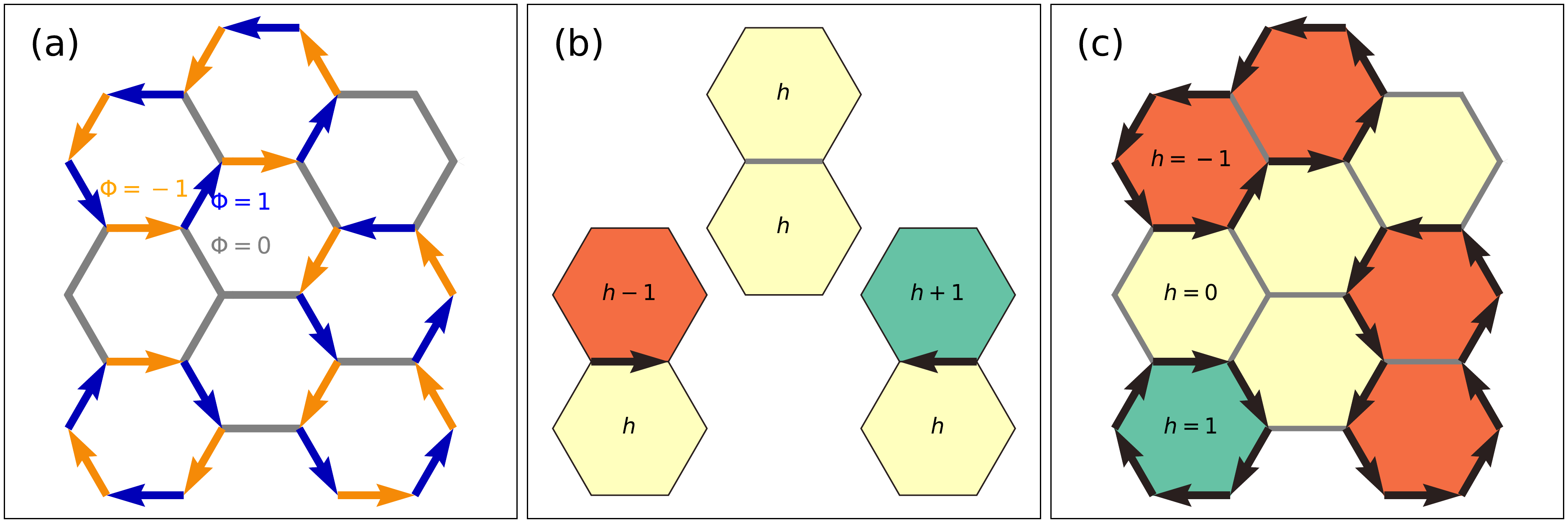}
	\caption{\textbf{Mapping on a topographic map.}
		\textbf{(a)} Configuration of our model (see Fig.~\ref{fig:example_spinmodel}).
		\textbf{(b)} Rules for the topographic map.
		If there is no arrow, the two faces have the same height. If there is an arrow pointing to the right, when moving from $f$ to $f'$, then $h_{f'} = h_f - 1$. If there is an arrow to the left, then $h_{f'} = h_f + 1$.
		\textbf{(c)} Topographic map corresponding to (a). In this configuration, $\Delta h = \max h - \min h = 2$.
	}
	\label{fig:height}
\end{figure}

In practice we measure the height field $h_f$ recursively.
The procedure is  easily understood from the sketch of the algorithm shown in Fig.~\ref{fig:height}a. 
In short, starting from a face $f$ associated with a height $h$ we move to the neighbouring face $f'$ by crossing an edge $e$.
If $e$  hosts a spin pointing towards the right (resp. left) hand side, we assign the value $h'=h-1$ (resp. $h'=h+1$). 
The right and left directions are defined with respect to the vector connecting the centers of $f$ and $f'$.
When the edge hosts no current ($\Phi_e=0$), then $h'=h$.
The resulting height field does not depend on the way the network is explored.
This procedure is very similar to the standard mapping of loop $O(n)$ models to solid-on-solid models discussed e.g. in Ref.~\cite{Nienhuis1982}.

The reason why we introduce this mapping is twofold. 
Firstly, the height amplitude $\Delta h = \max_\mcF h_f - \min_\mcF h_f$ is one of our key observables. 
It quantifies the nesting of the streamlines and distinguish between the two phases observed both in the experiments and in the simulations, see see section \ref{ss:compa}.
Secondly, this mapping will also allow us to map the streamlines of the crumpled phase to the domain walls of the antiferromagnetic Ising model on the triangular lattice of faces $\mcF$, see section~\ref{ss:nocoupling}. 


\section{Theory: Modeling the emergent flow patterns, 
	loops on the honeycomb lattice}

\subsection{Mapping active flows on a spin model} \label{ss:spinmodel}

\begin{figure}
	\centering
	\includegraphics[width=\textwidth]{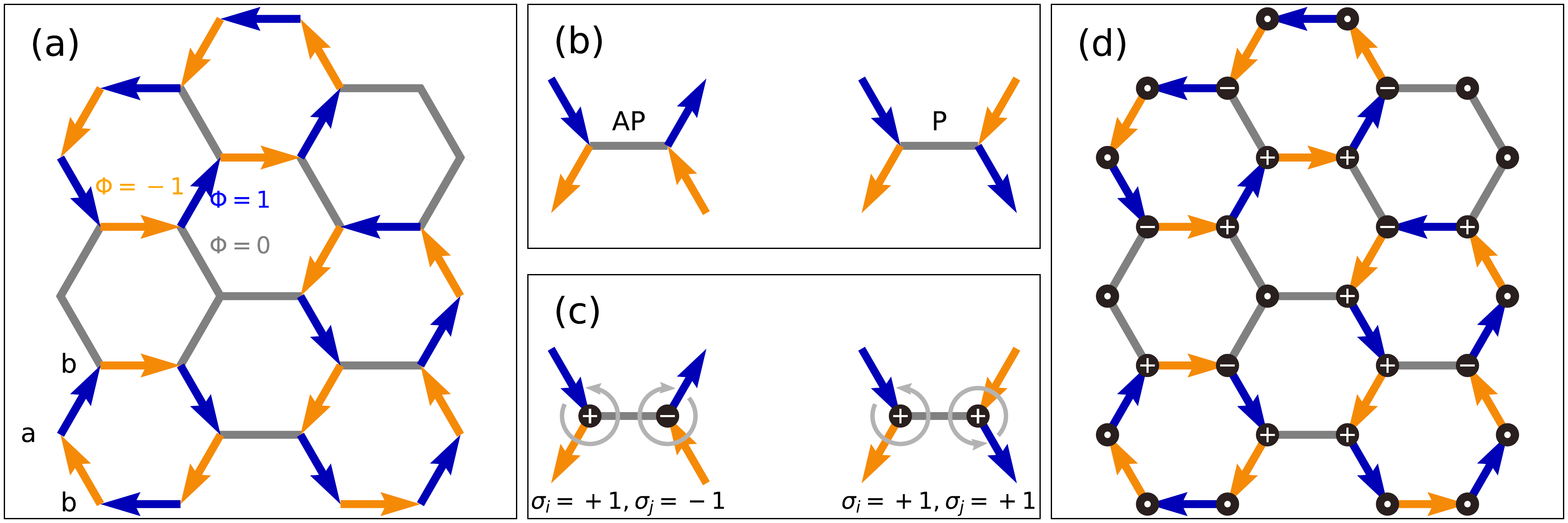}
	\caption{\textbf{Spin model.}
		\textbf{(a)} We consider a honeycomb network and recall that it is bipartite: `a'-vertices are adjacent to `b'-vertices only (see bottom left). We place spins $\Phi_e \in \{-1, 0, +1\}$ on the edges of the network with the local constraint $\sum\Phi_e = 0$ at each node (Eq.~\eqref{eq:massconservation}). $\Phi_e = +1$ corresponds to a flow from a `a'-vertex to a `b'-vertex  and is depicted as a blue edge while $\Phi_e=-1$ corresponds to the flow from a `b'-vertex to a `a'-vertex and is depicted as a orange edge. Finally, $\Phi_e=0$ corresponds to an absence of net flow and is shown in gray.
		\textbf{(b}) To account for the experimental flow patterns, we need to distinguish between antiparallel (AP) flows and parallel (P) flows across gray edges.
		\textbf{(c)} This is done by associating a spin $\sigma_i=+1$ to a vertex with (blue, orange, gray) edges in the counter-clockwise direction, and $\sigma_i=-1$ to a vertex with
		the same colors in the clockwise direction. An antiparallel coupling then corresponds to $\sigma_i\sigma_j = -1$, and a parallel coupling to $\sigma_i\sigma_j = 1$.
		\textbf{(d)} Same configuration as (a) with the vertex spins $+1$, $-1$ and $0$ shown as $+$, $-$ and $\bullet$.
	}
	\label{fig:example_spinmodel}
\end{figure}
In this section, we introduce a spin model that effectively accounts for the geometry of the streamlines observed in our active hydraulic experiments. 
Constructing a predictive model from first principle interactions between the active colloids, or even from the Toner-Tu hydrodynamic theory would be a formidable challenge. 
Here, we instead build a phenomenological description from conservation laws and our robust experimental observations.

\subsubsection{Honeycomb lattice geometry}
In all that follows, 
we note $\mcE$ the set of edges of the honeycomb network, the channels, and $\mcV$ the set of vertices, see Fig. \ref{fig:example_spinmodel}a.
Each edge links two vertices, and each vertex in the bulk connects to three edges.
In our numerical simulations, the networks are either be bounded, to make quantitative comparisons with our experiments (section \ref{ss:compa_expe}), or periodic to investigate the asymptotic scalings of the streamline geometry (section \ref{s:limits}).
It might be worth  noting that the edges of the honeycomb lattice form a Kagome lattice. 
We call $N$ the linear number of hexagonal faces ($N=3$ in Fig. \ref{fig:example_spinmodel}a), so that in a periodic system there are $N^2$ faces, $3N^2$ edges and $2N^2$ vertices.

Our model accounts for two constraints: (i) The active fluid current $\Phi_e$ flowing along the edge $e\in\mcE$ can only take three values in each channel; and (ii) the mass is conserved at every vertex.

\subsubsection{Active flows and Blume Capel spins}
To model condition (i), we consider Blume-Capel spins $\Phi_e \in \{-1, 0, 1\}$ \cite{Blume1971} on the edges $e=(i,j)\in\mcE$ of the honeycomb lattice 
(which are the vertices of the dual Kagome structure). 
$\Phi_e = 0$ corresponds to the absence of net flows along the edge $e$: the channel hosts vortices.  
$\Phi_e = \pm 1$ corresponds to  a finite current along  $e$.
The sign of $\Phi_e$ indicates the direction of the flow as follows.
The honeycomb lattice is bipartite: the vertices in $\mcV$ can be decomposed into two sublattices: `$\mcV_a$' and `$\mcV_b$', see Fig. \ref{fig:example_spinmodel}a. The edges only link `a'-vertices to `b'-vertices.
Given this observation, we assign $\Phi_e = +1$ to flows from an `a'-vertex to a `b'-vertex, and $\Phi_e = -1$ to flows from a `b'-vertex to an `a'-vertex. 
To distinguish between the three possible flow configurations, we associate a different color to each value of $\Phi_e$ as illustrated in Fig. \ref{fig:example_spinmodel}a: blue for $\Phi_e=+1$, orange for $\Phi_e=-1$, and gray $\Phi_e = 0$. 

\subsubsection{Mass conservation and spin-ice rule}
In steady state, mass conservation imposes a hard constraint on the Blume-Capel spin configurations.
The sum of the flows vanishes at each vertex, which translates into 
\begin{equation}
	\sum_{e\in\partial i} \Phi_e = 0
	\label{eq:massconservation}
\end{equation}
where $\partial i$ is the subset of edges connected to the vertex $i$.
In the bulk $\partial i$ includes three spins, associated with seven configurations: $\{+1, -1, 0\}=$ \{blue, orange, gray\} (and all possible permutations), or $\{0, 0, 0\}=$ \{gray, gray, gray\} as illustrated in  Fig. \ref{fig:example_spinmodel}a.
On the boundary of finite networks, there are two spins in $\partial i$, and they can only take the values $\{+1, -1\}$, $\{-1, +1\}$ or $\{0, 0\}$. 

We stress that the mass-conservation constraint of~\eqref{eq:massconservation} here corresponds to a spin-ice rule. 
As a matter of facts, in spin-ice models  frustration arises from a similar constraint: the sum of the magnetic moments at each site of a crystalline spin lattice must vanish~\cite{Moessner2006,Udagawa2021}.
We will show that the ice rule induces a strong degeneracy of the possible spin configurations in steady state.

\subsubsection{Self-avoiding loops}
As discussed in the main text, the spin-ice rule constrains the spin configurations, i.e. the streamlines of the active flow, to form oriented self-avoiding  loops on $\mcE$, Fig.~\ref{fig:example_spinmodel}a. 
This emergent geometry  can be understood as follows. 
We start from  an edge hosting a $+1$ spin (blue edge) pointing in the direction of the vertex $i$. 
The spin-ice rule imposes that 
this vertex is also connected to a single $-1$ spin (orange edge) pointing in the  direction of a neighboring vertex.
We can then iterate this reasoning to construct a walk on the honeycomb lattice. 
As each vertex can only connect to a $+1$ and a $-1$ edge the walks cannot intersect or cross themselves as these crossings would require vertices connected to three non-zero spins.
Finally, the resulting walks necessarily form closed loops in virtue of mass conservation. 

We note in passing that any ensemble of oriented self-avoiding loop on the honeycomb lattice can be tough as an ensemble of Blume-Capel spins on a Kagome lattice and obeying the spin ice rule. 
\subsubsection{Interactions between the loops}
As discussed in the main text, the spin-ice rule is not sufficient to explain the the geometry of the active hydraulic flows observed in our experiments. 
To make progress, we need to account for the ``parallel'' and ``antiparallel'' flow couplings across non-flowing edges, Fig.~\ref{fig:example_spinmodel}b. 
We therefore add a minimal ingredient to our spin model.
Fig.~\ref{fig:example_spinmodel}b would naturally suggest to include four-spin (anti)ferromagnetic interactions. 
Although possible this approach would lead to rather complex couplings.
In order to simplify our model, we instead assign an additional spin variable $\sigma_i \in\{+1, -1, 0\}$ to each vertex $i\in\mcV$ visually defined in Fig.~\ref{fig:example_spinmodel}c.
This spin variable describes the handedness of the $\Phi_e$ spin configurations around a vertex. 
In short, we measure the values of the $\Phi_e$ around the vertex $i$ in the counter-clockwise direction. 
$\sigma_i=1$ when the the edge spins take the values  $(+1, -1, 0)$ (ie. blue, orange, gray) up to a circular permutation.
$\sigma_i=-1$ when the edge spins take the value  $(-1, +1, 0)$ (ie. orange, blue, gray).
Finally $\sigma_i = 0$ at vertices where the three edge spins are $(0, 0, 0)$ or at the boundary of  finite systems.
Fig.~\ref{fig:example_spinmodel}c then readily tells us that (anti)parallel couplings between the streamlines translate into (anti)ferromagnetic interactions between the $\sigma$ spins.
We note that this construction is closely related to the color-dependent interactions of the 3-coloring model defined in Ref.~\cite{Verpoort2018}.
\subsubsection{Hamiltonian description of the stream lines}
We can now define a phenomenological  
Hamiltonian model to account for the experimental flow patterns.
We introduce the energy $\mathcal{H}$ of a spin configuration $\{\Phi_e\}$:
\begin{equation}
	\label{smeq:energyFull}
	\mathcal{H}(\{\Phi_e\}) =
	-J_A \sum_{e\in\mathcal{E}} \Phi_e^2
	-J_C \sum_{e=(i, j) \in \mathcal{E}} \delta_{\Phi_e, 0}
	\sigma_{i} \sigma_{j},
\end{equation}
where $\sigma_{i}$ and $\sigma_{j}$ are functions of $\{\Phi_e\}$ and $\delta_{\Phi_e, 0}=1$ if $\Phi_e=0$ and $0$ otherwise.
The two coupling constants $J_A$ and $J_C$ have a clear physical meaning.
$J_A > 0$ is a positive constant that promotes spontaneous flows, i.e that favors  finite spin values $\Phi_e = \pm 1$.
$J_C$ can be either positive, or negative, to account for the parallel, or antiparallel, couplings between streamlines separated by channels hosting vortices.



\subsection{Numerical methods: Monte-Carlo worm algorithm} \label{ss:monteCarlo}
\begin{figure}
	\centering
	\includegraphics[width=0.8\textwidth]{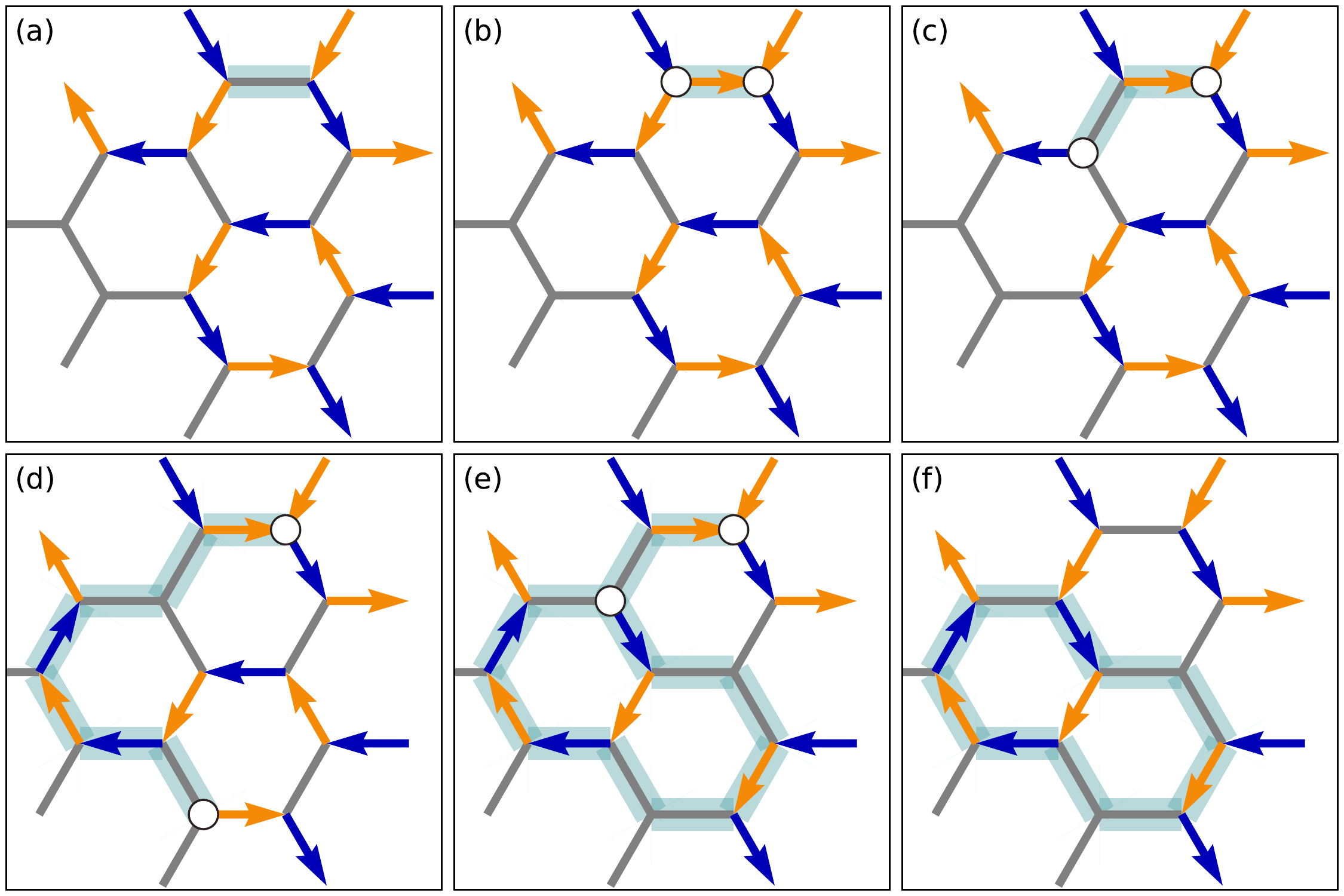}
	\caption{\textbf{Worm algorithm.} 
		\textbf{(a)} We select an edge in the initial configuration. 
		\textbf{(b)} Flipping the spin (here $0\mapsto -1$) leads to the creation of a pair of defects (white circles).
		\textbf{(c)} We move one of the defects by  flipping another edge spin.
		\textbf{(d-e)} We iterate the procedure until we come back to a site we already visited.
		\textbf{(f)} The beginning of the path is discarded. We use the Metropolis rule on the move that goes from configuration (a) to configuration (e).}
	\label{fig:worm_algorithm}
\end{figure}

Our idea consists in describing the  streamlines of the active hydraulic flows  as the low energy states of $\mathcal H$ satisfying the spin-ice constraint 
\eqref{eq:massconservation}.
To find them, we use a Monte-Carlo algorithm. 
We therefore define a   statistical mechanics model given by the following Boltzmann weights at inverse temperature $\beta$,
\begin{equation}\label{smeq:probaSpin}
	P(\{\Phi_e\}) = \frac{e^{-\beta \mathcal{H}(\{\Phi_e\})}}{Z},
\end{equation}
with the partition function
\begin{equation} \label{smeq:partitionFun}
	Z = \sum_{\{\Phi_e\} \in \mathrm{OL}(\mcE)} e^{-\beta \mathcal{H}(\{\Phi_e\})}.
\end{equation}
$\mathrm{OL}(\mcE)$ is the ensemble of 
oriented loops on the honeycomb lattice. 
Sampling the spin configurations on this ensemble guarantees that the spin-ice rule is satisfied.
In our numerical simulations, we set $J_A = 1$ without loss of generality and focus on the low temperature limit $\beta\gg 1$. 
The main parameter of our model is then $J_C$ that regulates the coupling of the flows at the two ends of a non-flowing channel.
To satisfy the spin-ice rule, we use a standard tool for lattice loop models, namely a worm Monte Carlo algorithm \cite{Barkema1998,Jaubert2011}. 
In brief, the algorithm consists first in generating
a worm, that is to say a loop of edges that can be flipped without violating the ice rule. The Monte Carlo move then corresponds to flipping the edge spins from $\pm 1$ to $0$ or $0$ to $\pm1$ along this worm. 
In details, 
we use an adaptation of the short loop algorithm developed in Ref.~\cite{Barkema1998} for the three-color model. 
We initialize the simulation with all edge spins equal to zero ($\Phi_e=0$).
This configuration trivially satisfies the ice rule. 
We then generate a new allowed configuration by constructing a worm. 
We explain the procedure to generate it in general, and not only for the initial configuration.
Starting from a given spin configuration $\{\Phi_e\}$ satisfying the ice rule, we choose an edge $e$ at random, Fig.~\ref{fig:worm_algorithm}a.
It is the first edge of the worm.
If $\Phi_e = \pm 1$, we update it to $\Phi_e^\text{new} = 0$, and if $\Phi_e=0$ we update it at random to $\Phi_e^\text{new}=\pm1$, Fig.~\ref{fig:worm_algorithm}b.
In both cases, this creates a pair of defects to the ice rule on the two ends of the edge $e=({i,j})$: these defects are referred to as charges in spin ice models: $Q_i = \sum_{e\in \partial i} \Phi_e \neq 0$.
To construct the second edge of the worm, the idea is then to propagate the charges on the lattice until they annihilate. 
To do so, we consider the site $i$ and try to recover $Q_i = 0$ by changing the spin of one of the two neighboring edges of $i$ (different from $e$), Fig.~\ref{fig:worm_algorithm}c.
When a single move is possible, we accept it and the second edge of the worm is defined.
If there exists two possible moves, we choose one at random, which then defines the second edge of the worm.
We 
iterate this procedure which results in the construction of a 
walk on the honeycomb lattice, Fig.~\ref{fig:worm_algorithm}d. 
When this walk visits a site $k$ that has already been visited, we form a self-avoiding loop attached to a dangling end at $k$, Fig.~\ref{fig:worm_algorithm}e.
We remove the dangling end by discarding all the spin flips used to construct the 
first steps of the random walk, Fig.~\ref{fig:worm_algorithm}e.
This procedure is known as the short loop algorithm~\cite{Barkema1998}.

We can now evolve the Monte-Carlo dynamics by computing the energy from Eq.~\eqref{smeq:energyFull}.
The spin-flips along the worm are then accepted according to the standard Metropolis rule~\cite{Krauth2006} using the Boltzmann weight \eqref{smeq:probaSpin} at inverse temperature $\beta$.
This algorithm is ergodic (any configuration can be reached from the all-zero state and vice-versa) and satisfies detailed balance (Metropolis rule): it thus samples the ensemble of configurations compatible with the ice rule according to the Boltzmann probability distribution \eqref{smeq:probaSpin}.

We performed two types of simulations. (1) Small systems without periodic boundary conditions to compare our numerical results to our measurements. (2) Large systems with periodic boundary conditions in order to compute accurate scaling exponents.
Simulations (1) are performed at system size $N=30$, inverse temperature $\beta=2$, flow constant $J_A = 1$ and coupling constant $J_C = -2, -1, -0.5, 0, 0.5, 1, 2$. 
For each set of parameters, we perform  $50$ independent simulations.
We perform $10^6$ worm iterations to thermalize the system, and then $10^6$ additional iterations during which we compute the observables every $10^3$ iterations. We compare our  results with our experiments in Section~\ref{ss:compa_expe}.
Simulations (2) are performed for system sizes ranging from $N=20$ to $N=500$, $\beta=1$, $J_A=1$ and $J_C=-1, 0, 1$. 
For each set of parameters, we perform $10$ independent simulations, with $20000N^2$ thermalization iterations and $20000N^2$ more iterations during which the observables are computed every $200N^2$ iterations. 
We report our results in Section~\ref{s:limits}.

\begin{figure}
	\centering
	\includegraphics[width=\textwidth]{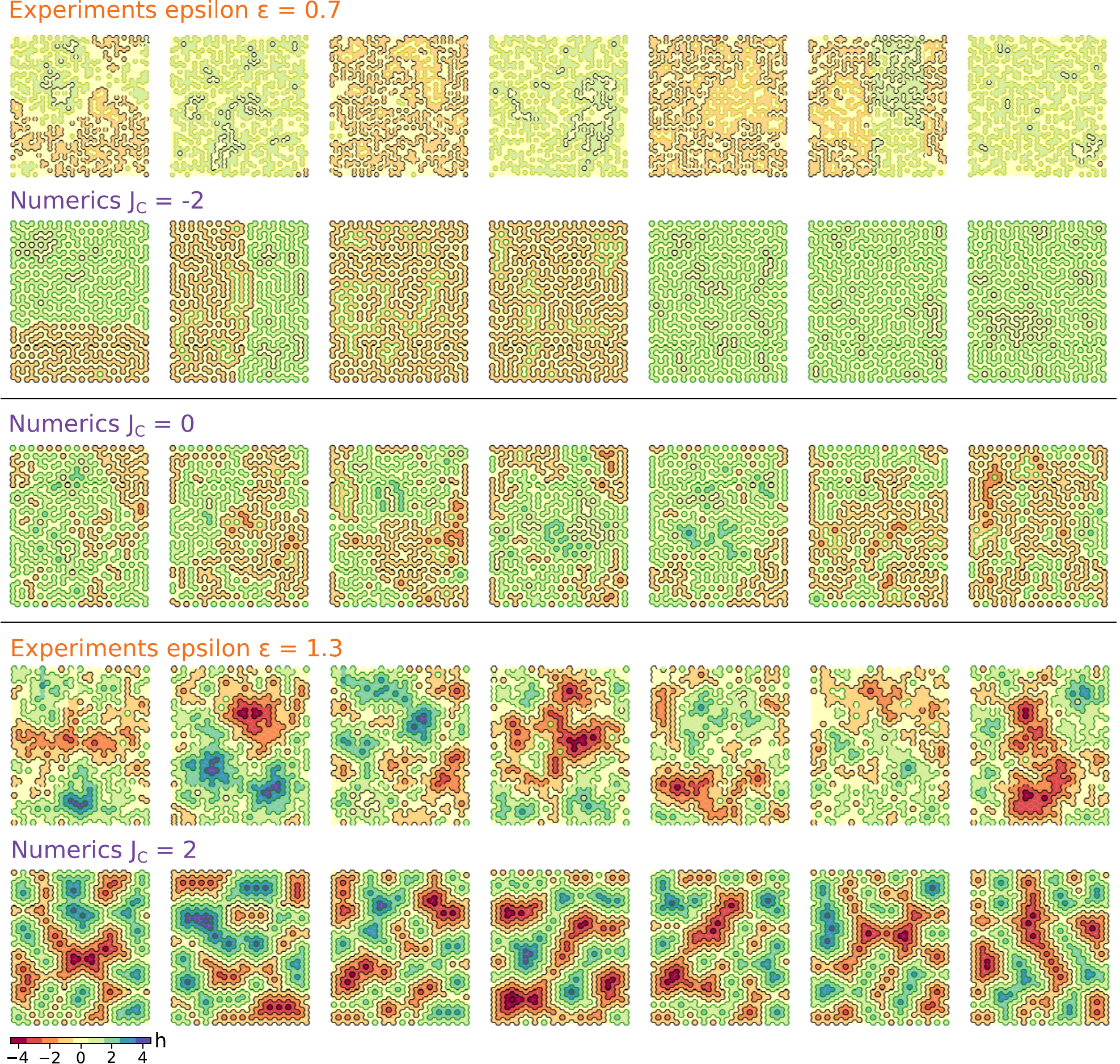}
	\caption{\textbf{Degeneracy of the flow patterns.} Counter-clockwise loops are shown in black, and clockwise loops in green. The colors of the faces correspond to their heights (see subsection \ref{ss:height} and Fig. \ref{fig:example_spinmodel}c).
		Top: antiparallel regime with experimental configurations at $\epsilon=0.7$ and numerical configurations at $J_C = -2$.
		Middle: numerical configurations with no flow coupling ($J_C=0$).
		Bottom: parallel regime with experiments at $\epsilon=1.3$ and numerics at $J_C = 2$.
	}
	\label{fig:degeneracy}
\end{figure}

\begin{figure}
	\centering
	\includegraphics[width=0.9\textwidth]{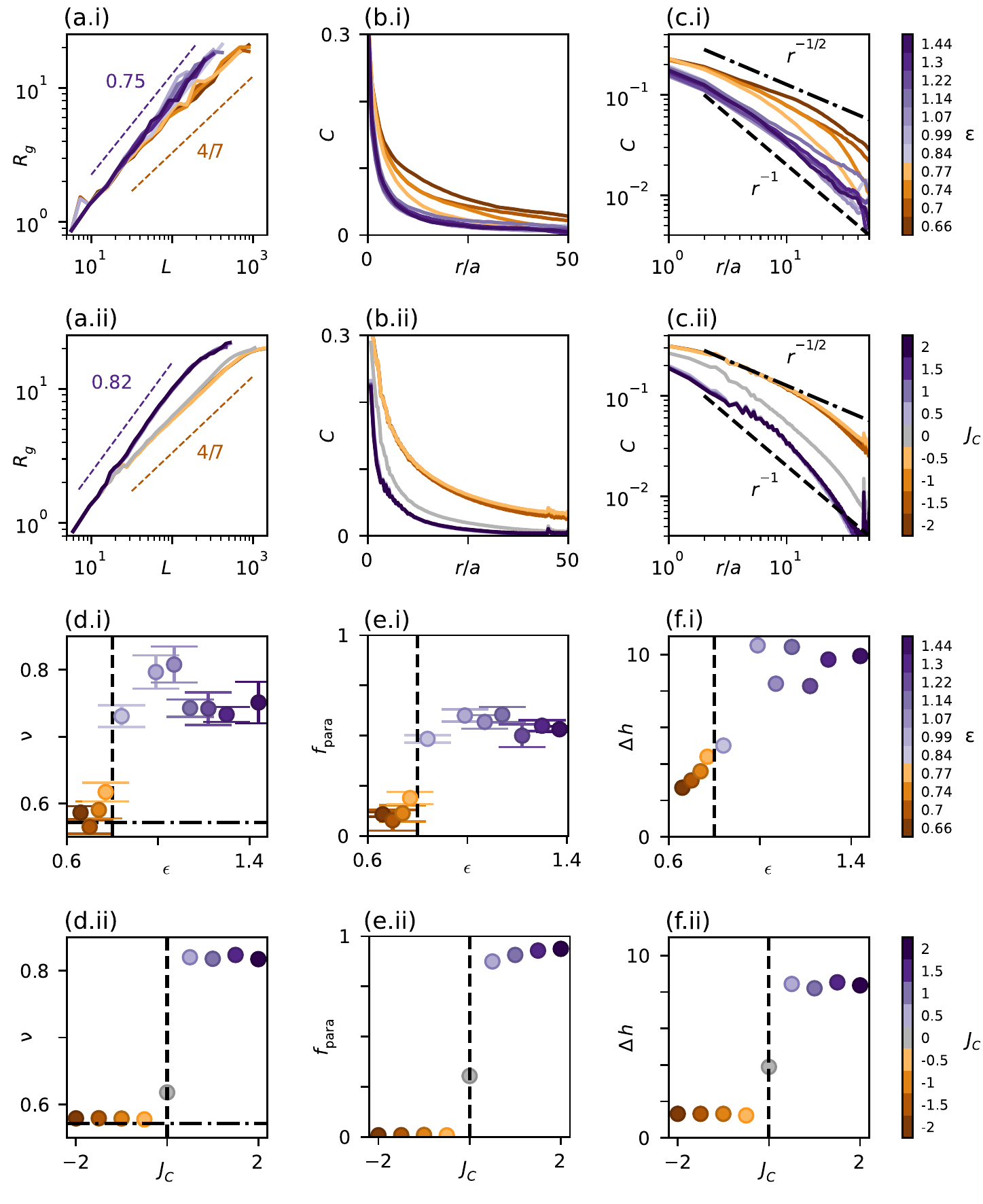}
	\caption{\textbf{Quantitative comparison between numerical and experimental results.}
		\textbf{(a.i)} and \textbf{(a.ii)} Average gyration radius of a loop as a function of its length for experimental aspect ratio $\epsilon \in [0.55, 1.44]$ (in (a.i)) and numerical coupling constant $J_c\in[-2, 2]$ (in (a.ii)).
		%
		\textbf{(b.i)} and \textbf{(b.ii)} Experimental and numerical probability $C(r)$ that two sites at a distance $r$ belong to the same loop.
		\textbf{(c.i)} and \textbf{(c.ii)} Same curves in log-log scale. The dashed and dash-dotted lines respectively correspond to exponents $-1$ and $-1/2$.
		\textbf{(d.i)} and \textbf{(d.ii)} Exponent $\nu$ of the gyration radius, fitted from (a.i) and (a.ii). The error bars are obtained from the least-square fit algorithm with input uncertaintly given by the spread in each bin of (a). 
		The horizontal dash-dotted line corresponds to the prediction of the antiferromagnetic Ising model ($4/7$, see subsection \ref{ss:antiferro}).
		\textbf{(e.i)} and \textbf{(e.ii)} Fraction $f_\mathrm{para}$ or parallel streamlines across non-flowing edges.
		\textbf{(f.i)} and \textbf{(f.ii)} Height amplitude $\Delta h$ (see Fig. \ref{fig:example_spinmodel}c).
	}
	\label{fig:compa_expe}
\end{figure}

\section{Comparison between simulations and experiments} \label{ss:compa_expe}
\subsection{Qualitative features}
We performed simulations of our spin model with $30\times 30$ hexagons, which is close to the size of our microfluidic networks (see section \ref{ss:monteCarlo}). 
In the experiments, the control parameter is the aspect ratio $\epsilon$ of the channels.
It controls the fraction of antiferromagnetic couplings.
For $\epsilon<\epsilon^\ast=0.8$ the flow couplings are mostly antiparallel while for $\epsilon>\epsilon^\ast$ they are mostly parallel. 
The analog in the simulations is the coupling constant $J_C$:  $J_C < 0$  favors antiparallel couplings, whereas  $J_C > 0$ promotes parallel couplings. 

Fig. \ref{fig:degeneracy} confirms that  the sign $J_C$ is a very good proxy for the role of the channel aspect ratio of the streamline geometry.
We indeed find that when  $J_c < 0$ the spin loops are crumpled and collapsed, as the streamlines in the experiments where $\epsilon<\epsilon^\star$.
In addition, the equivalent topographic map typically includes only two to three  values, again in excellent agreement with our experimental findings. 
In short channels, the streamline soup does not form a nested structure.
When  $J_c>0$, we find that the spin loops are more persistent and form nested structures, the height functions takes a range of values spanning a typical interval $\Delta h \simeq 8$ for our system sizes. 
These geometrical feature again reproduce what we observe in our experiments where the streamlines are persistent and crumpled when $\epsilon>\epsilon^\star$.
In all cases our simulations also reflect the strong degeneration of the steady states observed in our active hydraulic networks.

This qualitative agreement begs us to quantitatively characterize the geometries of the two loops networks to validate the relevance of our analogy between frustrated Blume-Capel spins and active hydraulic flows

\subsection{Quantitative comparison: geometry of the streamlines and spin loops} \label{ss:compa}

To be more quantitative we consider all the observables accessible  to our experiments and proceed to a systematic comparison with our simulations,  Fig. \ref{fig:compa_expe}.

\begin{itemize}
	\item \textbf{Fraction of parallel contacts.} An essential observable  distinguishes the two active flow patterns: the fraction of parallel contacts across non-flowing edges $f_\mathrm{para} = n_\mathrm{para}/(n_\mathrm{para}+n_\mathrm{anti-para})$ where $n_\mathrm{para}$ is the number of  edges hosting no net flow ($\Phi_e = 0$ in the simulations) and separating two streamlines flowing in the same direction ($\sigma_i\sigma_j > 0$ in the simulations), $n_\mathrm{anti-para}$ counts the edges separating streamlines flowing in opposite directions.
	In the experiments, $f_\mathrm{para}$ varies from $\sim 0.1$ for $\epsilon<\epsilon^\ast$ to $\sim 0.6$ for $\epsilon<\epsilon^\ast$, see Fig. \ref{fig:compa_expe}e.i.
	Anti-parallel contacts dominate in the phase where the loops are crumpled while they are subdominant in the phase where the loops are more persistent and nested. This distinction is even more marked in the simulations: $f_\mathrm{para}$ is very close to $0$ for $J_c < 0$ while it reaches $0.9$ when $J_c > 0$ (Fig. \ref{fig:compa_expe}e.ii).
	
	\item \textbf{Gyration radius.} To distinguish between the crumpled loops of the first regime and the more swollen loops of the second regime, we compute the gyration radius $R_g$ of each loop,
	\begin{equation}
		R_g^2 = \frac{1}{L}\sum_e \left(\vec{X}_e - \frac{1}{L}  \sum_e \vec{X}_e\right)^2,
	\end{equation}
	where the sum runs over the $L$ edges $e$, at positions $\vec{X}_e$ that make the loop. 
	This is a typical observable for polymers~\cite{Doi1988} or spin-ice models~\cite{Jaubert2011}. We consistently find that the loops have a self-similar geometry, the average gyration radius scales as a power-law of the loop size,
	\begin{equation}
		\langle R_g\rangle \sim L^\nu.
	\end{equation}
	$\nu = 1/2$ corresponds to Brownian motion, $\nu = 1$ corresponds to straight lines or circles and $\nu = 3/4$ corresponds to self-avoiding walks in 2D. Note that $\nu= 1/d_f$ where $d_f$ is the fractal dimension of the model.
	
	In Fig. \ref{fig:compa_expe}a.i and a.ii, we plot the gyration radius of the loops. Our experiments and simulations are consistent. Loops in the parallel regime have a gyration radius that grows faster than those in the anti-parallel regime. 
	The curves  collapse on two master curves in the two distinct regimes. 
	We plot the variations of the exponent $\nu$ in Figs. \ref{fig:compa_expe}d.i and d.ii as a function of $\epsilon$ and $J_C$ respectively. 
	The transition is sharp both in the simulations and the experiments. 
	The anti-parallel regime corresponds to an exponent $\nu$ that is consistent with $4/7\simeq 0.571$. This exponent correspond to the gyration radius of the domain walls in   the anti-ferromagnetic Ising model, see subsection \ref{ss:antiferro}.
	In the parallel regime we find  a  higher exponent, around $0.75-0.8$ in the experiments and $0.82$ in the simulations. 
	We will argue below that there exist no indication for universality in this regime.
	
	\item \textbf{Loop correlation.} A typical observable that distinguishes loop models \cite{Peled2019,Jaubert2011} is the probability $C(r)$ that two edges separated by a distance $r$ belong to the same loop. 
	$C(r)$ decays as a power-law in the  critical phases of loop models.
	We plot this quantity in Fig. \ref{fig:compa_expe}b.i and b.ii. We see that both in the experiments and the simulations, $C(r)$ decays  faster in the anti-parallel phase than in the parallel phase.
	The system size does not allow us to conclude on a power-law decay and on its exponent. 
	However, the data in the anti-parallel regime are consistent with $C(r)\sim r^{-1/2}$ predicted for the anti-ferromagnetic Ising model (see section \ref{ss:antiferro}).
	Moreover, the data in the parallel regime may be consistent with $C(r)\sim r^{-1}$ (see section~\ref{ss:para}).  
	
	\item \textbf{Average height amplitude.} Finally, we quantify the nesting level of the loops and plot the height amplitude $\Delta h$ of the patterns in Figs. \ref{fig:compa_expe}f.i and f.ii. 
	We see that $\Delta h$  undergoes a sharp jump at the point where the fraction of parallel contact jumps. 
	This observation  holds both in our experiments and simulations, as anticipated from the patterns plotted in Fig. \ref{fig:degeneracy}.
\end{itemize}

This ensemble of results confirm that our minimal spin model capture all the salient features of the active hydraulic flows measured in our experiments.
The size currently accessible to our experiments does not allow to predict whether the transition between the crumpled and the nested streamline regimes is a genuine dynamical phase transition or a smooth crossover.
Answering this question goes beyond the scope of this article. However, we can take advantage of our numerical simulations to predict the asymptotic scaling laws that characterize the self-similar shape of the interacting self-avoiding loops.

\section{Asymptotic limits of the spin model and streamline geometry} \label{s:limits}
In the previous section, we introduced an equilibrium spin model (Eqs. \eqref{smeq:energyFull}-\eqref{smeq:partitionFun}) to account for the flow patterns observed in the experiments. 
We found a good agreement between experiments and simulations in both regimes of antiparallel and parallel flow couplings, whcih respectively correspond to small and large channel aspect ratio. 
To gain a more quantitative insight into the morphology of the  streamlines, we investigate numerically the two asymptotic limits: 
$J_C\to -\infty$ and $J_C\to +\infty$ (in the so called fully-packed limit where $J_A\to\infty$). 
We use theoretical arguments  and extensive numerical simulations to predict the scaling laws of the gyration radius  $\nu$ and of the decay of the correlation $C(r)$.
Throughout this section, all simulations are performed with  periodic boundary conditions.

This technical section is organized as follows: Section~\ref{ss:loopOn} is a reminder of the loop O($n$) models and of their mapping on spin models. Section \ref{ss:antiferro} focuses on the antiparallel regime ($J_C\to-\infty$). We show that this regime maps on the antiferromagnetic Ising model on a triangular lattice, or equivalently the loop O($1$) model. This mapping allows us to predict the  gyration radius exponent ($\nu=7/4$) and the algebraic decays $C(r)\sim r^{-1/2}$. 
In Section \ref{ss:nocoupling}, we investigate the case where there is no orientational coupling between the streamlines ($J_C=0$). 
While this situation cannot be 
realized in our experiments, it provides an insight into what happens in the crossover between the two asymptotic regimes ($J_C\ll0$ and $J_C\gg0$). 
When $J_C=0$ our model maps on a three-color model, or equivalently to a loop O($2$) model. 
In this regime we predict a gyration radius exponent $\nu = 2/3$ and a correlation decaying as $C(r)\sim r^{-1}$ consitent with our experimental findings at intermediate aspect ratio. 
Finally, in Section \ref{ss:para}, we discuss  the parallel regime ($J_C\gg0$). 
This regime is not as simple as  the two others as the ground state of our model features a sub-extensive degeneracy. 
The simulations reach metastable states which we characterize numerically.

\subsection{Mapping spin statistics on loop O($n$) models} \label{ss:loopOn}
The spin configurations constrained by the spin-ice rule correspond to oriented loops on the honeycomb lattice. 
Loop models have a long history in equilibrium statistical mechanics. A classic class of loop models is the loop O($n$) models \cite{Peled2019,DuminilCopin2020}.
We briefly review some key results as this class of model offer powerful tools to explain the asymptotic limits of our spin model, and the corresponding active-flow geometry.

\medskip
\textbf{Definition of the loop O($n$) model.} Let us consider the set of edges $\mcE$ of a finite lattice, in our case the honeycomb lattice. 
We define $L(\mcE)$, the set of all loop configurations on $\mcE$. 
A loop configuration $\omega$ is a subset of $\mcE$ for which all the vertices of the underlying graph have an even connectivity: in our case each vertex is connected to either $0$ or $2$ edges. 
We denote $\#\mathrm{edges}(\omega)$ the number of edges in a loop configuration $\omega$, and $\#\mathrm{loops}(\omega)$ the number of loops. 
By definition, the probability of $\omega$ in the loop O($n$) model is given by
\begin{equation}
	P^\mathrm{loop}(\omega) = \frac{x^{\#\mathrm{edges}(\omega)} n^{\#\mathrm{loops}(\omega)}}{Z^\mathrm{loop}(x, n)}
\end{equation}
with the partition function
\begin{equation}
	Z^\mathrm{loop}(x, n) = \sum_{\omega\in L(\mcE)}x^{\#\mathrm{edges}(\omega)} n^{\#\mathrm{loops}(\omega)}.
\end{equation}
where $x$ is the fugacity of the number of edges and $n$ is the fugacity of the number of loops. 
The limit $x\to\infty$ corresponds to the fully-packed regime: the number of edges in the loop configuration is maximized.

\medskip
\textbf{Three important cases: $n=0$, $n=1$ and $n=2$.} 
It is known that the loop O($n$) model \cite{Peled2019} maps on canonical statistical mechanics models. 
The limit $n\to0$ corresponds to the celebrated De Gennes mapping on the statistics of self-avoiding random walks. 
The case $n=1$ maps on the Ising model on the triangular lattice of faces $\mcF$, with $x=e^{-2\beta}$ where $\beta$ is the inverse temperature of the Ising model. 
We will detail further and use this equivalence in Section \ref{ss:nocoupling}.
Finally, $n=2$  at $x=\infty$ is equivalent to a three-coloring model which we discuss in Section \ref{ss:antiferro} to investigate the $J_C=0$ phase.

\medskip
\textbf{Relation between spin and loop O($n$) models.} 
We recall that for integer $n$ the loop O($n$) model is closely related to the spin O($n$) model \cite{Blote1994,Peled2019}. 
Considering $n$-dimensional spins $\sss_i = (s_i^1, \dots, s_i^n)$, with $\sum_k s_i^k = n$, on the vertices $i\in\mcV$ of the graph, the partition function of the spin O($n$) model reduces to 
\begin{equation}
	Z^\mathrm{spin}(x, n) = \int d\sss_1 \dots d\sss_N \prod_{(i, j)\in \mcE} w(\sss_i\cdot\sss_j).
\end{equation}
Writing the weight function as $w(\sss_i\cdot\sss_j) = 1+x\sss_i\cdot\sss_j$, one can show that $Z^\mathrm{spin}(x, n) = Z^\mathrm{loop}(x, n)$ \cite{Blote1994,Peled2019}.

\medskip
\textbf{Criticality and Schramm-Loewner evolution.}
One important question for the loop O($n$) model, for a given loop fugacity $x$, is whether the loops have a characteristic size or not.
In other words, does the probability of finding a loop of length $L$, $P_\mathrm{loop}(L)$, decay exponentially or algebraically?
This question was answered by  Nienhuis in~\cite{Nienhuis1982}.
He showed that for $n\in [0, 2]$ there exists a critical point at $x_c(n) = \left[2+(2-n)^{1/2}\right]^{-1/2}$: when $x<x_c$ $P_\mathrm{loop}(L)$ decays exponentially (sub-critical phase), whereas for $x> x_c$ and $x=x_c$ $P_\mathrm{loop}(L)$ decays as a power-law (critical phase). 
When $x>x_c$, 
the loops are non only scale-invariant but  conformaly invariant.
It was then conjectured in \cite{Kager2004} than in this critical phase, the scaling of the loop size  is given a Schramm-Loewner evolution $\mathrm{SLE}(\kappa)$ \cite{Cardy2005} with parameter $\kappa\in[4, 8]$ given by
\begin{equation}
	n = -2\cos(4\pi / \kappa).
\end{equation}
In other words $\kappa=8$ for $n=0$, $\kappa=6$ for $n=1$, and $\kappa=4$ for $n=2$. 
The technical aspects of these results go beyond the scope of our article. 
We simply recall that a Schramm-Loewner curve of parameter $\kappa$ \cite{Cardy2005} corresponds to a stochastic evolution in the half-plane having a fractal dimension $d_f$ and that features conformal invariance, 
\begin{align}
	d_f &= 1 + \frac{\kappa}{8}. 
\end{align}
In particular $d_f = 7/4$ 
for $\kappa=6$ ($n=1$);  $d_f=3/2$ 
for $\kappa=4$ ($n=2$).
The  correspondence between loop O($n$) models and Schramm-Loewner evolutions (or more precisely conformal loop ensembles) is the topic of numerous works in the mathematical litterature but rigorous results remain scarce~\cite{Peled2019,DuminilCopin2020}.

In the following two sections, we show that the phases $J_C < 0$ and $J_C=0$ of our model correspond to fully packed loop O($1$) and O($2$) models.

\subsection{Antiparallel flow couplings: antiferromagnetic Ising model on the triangular lattice} \label{ss:antiferro}
In this subsection, we investigate our model (Eqs. \eqref{smeq:energyFull}-\eqref{smeq:partitionFun}) in the limit $J_C \ll 0$, i.e  
in the limit where all   adjacent streamlines flow along opposite directions. 
This limit is directly relevant for  experiments where the channel aspect ratio is small.

\subsubsection{Mapping of our model on the antiferromagnetic Ising model in the limit $J_C\to -\infty$}
We
focus on the ground states, $\beta\to\infty$, of our spin model when $J_C < 0$, with $J_A=1$. 
We show that they map on the ground states of the antiferromagnetic Ising model on the triangular lattice.

We first note that the constraint of maximal flow $\beta J_A\to\infty$ implies that all vertices have their three edges associated with spins $\{+1, -1, 0\}$, there are no $\{0, 0, 0\}$ vertices. 
This is readily seen by decomposing the sum 
$\sum_{e\in\mcE} \Phi_e^2 = \frac{1}{3} \sum_{i\in\mcV} \sum_{e\in\partial i} \Phi_e^2$ where $\partial i$ is the set of the three edges adjacent to $i$. This sum is maximized if all the internal sums are equal to $2$, hence all vertices must have a $\{+1, -1, 0\}$ structure.

\begin{figure}
	\centering
	\includegraphics[width=\textwidth]{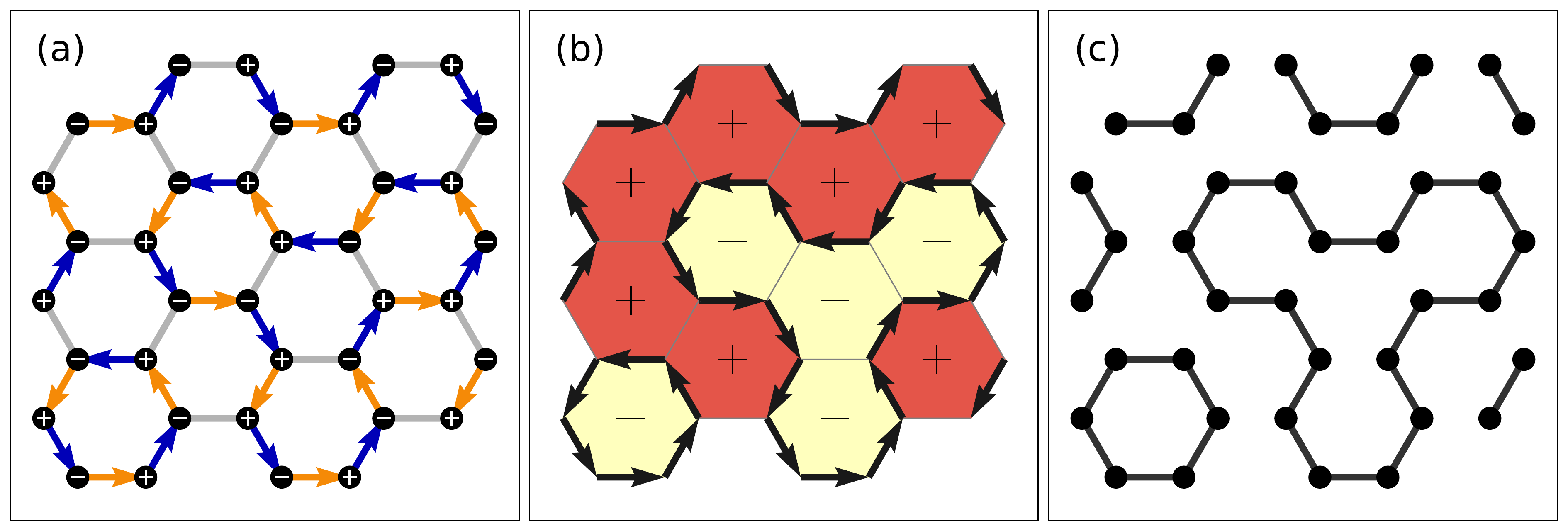}
	\caption{\textbf{Antiparallel regime ($J_C < 0$): Mapping our active flow model on the antiferromagnetic Ising model.}
		\textbf{(a)} Example of fully-packed configuration of our spin model (Eq. \eqref{smeq:energyFull}) with antiparallel couplings only ($\sigma_i\sigma_j = -1$ on all zero edges). 
		The spin arrows reflect the local direction of the active flow  in the fluidic netowrk
		\textbf{(b)} Mapping of the spin configuration of (a) on a topographic map (see Section \ref{ss:height}). 
		We note that the height field can only take two different values, which we associate to two Ising spins $+$ and $-$ to them. 
		In this limit, the streamlines of the streamlines of the active hydraulic flows map on the domain walls of an antiferromagnetic Ising model.
		\textbf{(c)} Another perspective on our statistical mechanics problem consists in looking at the shape of the closed streamlines irrespective of their orientation (dark lines). They  define a realization of a loop O($1$) model.
	}
	\label{fig:example_isingaf}
\end{figure}

Second, the limit $\beta J_C\to - \infty$ implies $\sigma_i\sigma_j = -1$ whenever a non-flowing edge ($\Phi_e = 0$) connects the sites $i$ and $j$.
Adjacent stream lines flow in opposite directions.
Constructing the topographic map of subsection \ref{ss:height}, we then readily find that the height field can only take the values $h=0$ and $h=+1$, or $h=0$ and $h=-1$, see Fig. \ref{fig:example_isingaf}.
We can make yet again another spin analogy. For each face $f\in\mcF$, we associate the two possible height values to Ising spins $S_f=+1$ when $h=0$ and $S_f=-1$ when $h=+ 1$.

Finally we can show that the limit $\beta J_C\to - \infty$ implies antiferromagnetic couplings between the $S_f$ spins. 
This is done by noticing that if $f$ and $f'$ are the two faces in contact with an edge $e$, then we can relate the edge-spin value to the face-spin values as 
$\Phi_e^2 = (1-S_fS_{f'})/2$. This equality follows from  the rule at the origin of the construction of the topographic map.
Using this simple relation we can recast the term that promote active flows on each edge as:
\begin{equation}
	-J_A \sum_{e\in\mcE} \Phi_e^2 = \frac{J_A}{2} \sum_{f\sim f'} S_f S_{f'} -\frac{J_A N_\mathrm{edges}}{2},
\end{equation}
which is nothing else but the energy of an antiferromagnetic Ising model
\begin{equation}
	E_\mathrm{Ising} = J_\mathrm{Ising}\sum_{f\sim f'} S_f S_{f'} + \mathrm{const.}
\end{equation}
where $J_\mathrm{Ising} = J_A / 2 > 0$. 
In the limit $\beta J_A\to+\infty$ the geometry of the streamlines therefore correspond to geometry of the domain walls in the ground state of an antiferromagnetic Ising model on the triangular lattice.
This mapping readily explains the extensive degeneracy of the ground states which is a well established result for antiferromagnets on the triangular lattice, see e.g. \cite{Blote1993}.

In the next section we take advantage on this last frustrated-magnetism analogy to make exact predictions about the streamline geometries of active flows in trivalent hydraulic networks.

\begin{figure}
	\centering
	\includegraphics[width=\textwidth]{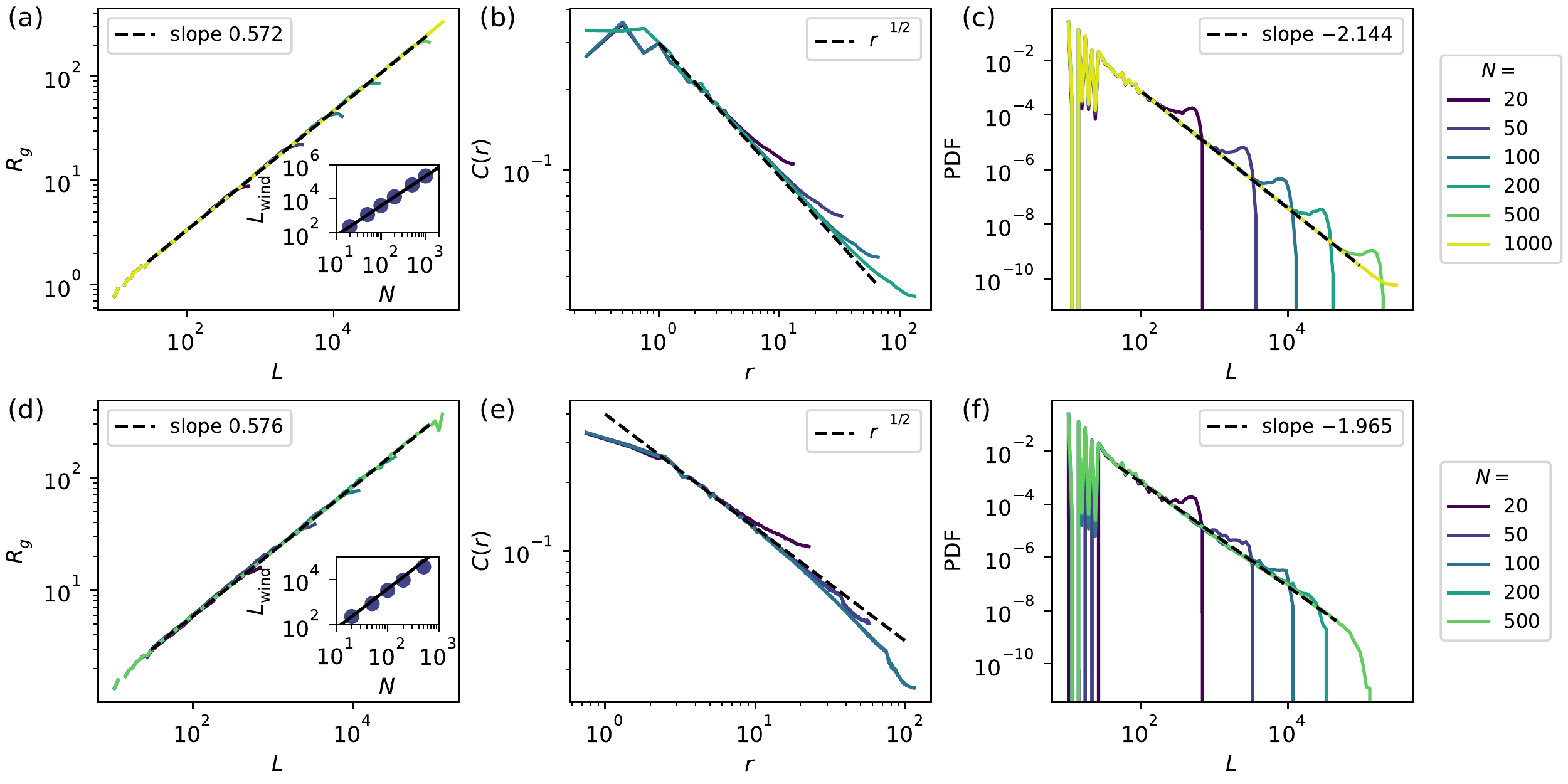}
	\caption{\textbf{Antiparallel regime ($J_C < 0$): numerical results.}
		\textbf{(a-c)} Monte-Carlo simulations of the Ising model. \textbf{(d-f)} Monte-Carlo simulations of our spin model with $J_C=-2$. 
		\textbf{(a)} and \textbf{(d)} show the gyration radius $R_g$ as a function of loop size $L$ for system sizes $N=20$ to $1000$, excluding  loops which wind around the periodic boundaries.
		The exponents of the power laws found in the two sets of simulations are respectively $0.572$ and $0.576$ in agreement with the exact prediction $\nu = 4/7 \approx 0.573$.
		In the inset, the average lengths of the loops that wind around the system is plotted as a function of the system size. The growth is in agreement with $N^{7/4}$ (black line) which is the fractal dimension $d_f=1/\nu$.
		\textbf{(b)} and \textbf{(e)} show the decay of the probability $C(r)$ that two edges at a distance $r$ belong to the same loop. 
		The two sets of simulations are consistent with an algebraic decay as  $r^{-1/2}$.
		\textbf{(c)} and \textbf{(f)} Distribution of loop sizes $L$ (excluding winding loops) for the two types of simulations. The prediction for the exponent is $\tau = 2+1/7 \approx 2.143$.
		For all three observables, the agreement with theoretical predictions  is more accurate for the Ising simulations. The statistics and the equilibration of the system are both better in this simulation set.
	}
	\label{fig:limit_antipara}
\end{figure}

\subsubsection{Geometry of active-hydraulics streamlines with antiparalle couplings: Exact results}
We have already shown that our experimental measurements are correctly accounted for by  our active hydraulic model (Eq.~\eqref{smeq:energyFull}) in finite-size systems.

We are now equipped to make exact predictions about the streamline geometry for asymptotically large system size, when anti-parallel couplings prevail. 
These exact predictions rely on the series of mappings discussed in the previous sections and on technical results from the mathematical physics literature, which we have 
connected to get at quantitative predictions:
(i) Our active hydraulic model maps on the antiferromagntic Ising model in the fully packed limit when $J_C\ll0$.
(ii) The ground state of the antiferromagnetic Ising model \cite{Blote1993}, is equivalently described by the ground state of the fully packed loop O($1$) model~\cite{Kondev1996}.
This model corresponds to a conformal field theory with central charge $c=1$. 
iii) The fully packed loop O($1$) is commonly believed to obey a $\mathrm{SLE}(\kappa=6)$ scaling \cite{Kager2004} (see Section \ref{ss:loopOn})\footnote{Note that the convergence to $\mathrm{SLE}(\kappa=6)$ has been proven rigorously for the critical percolation, which corresponds to the Ising model in the opposite limit of high temperature \cite{Smirnov2001}.}. We refer the reader to the more recent Refs. \cite{Jaubert2011,Kassel2016,Verpoort2018,Peled2019} for in-depth bibliography, and note that all our observations fall in the same universality class as the spin-ice loop model of Ref. \cite{Jaubert2011}.

In order to confirm that our reasoning is correct we confront all exact predictions with two sets of simulations results: the simulation of our active-hydraulics model and simulations of the low temperature phase of the antiferromagnetic model on the triangular model (we detail the simulation method in the next section).
We find an excellent agreement between our exact predictions and our numerical results.
They are consistent with all our experimental findings, thereby confirming our full understanding of active hydraulics in the anti-parallel regime.

\begin{itemize}
	\item \textbf{Gyration radius.}  The exponent $\nu$ of the gyration radius $R_g$ is the inverse of the fractal dimension $d_f=1/\nu$ of the Ising domain walls.
	The theoretical prediction for the ground state of the antiferromagnetic Ising model is $\nu=4/7\simeq 0.571$ ($d_f=7/4$). This result is established both using the SLE equivalence, and  the loop O($1$) equivalence~\cite{Kondev1995,Kondev1996}. 
	In our simulations of  the Ising model and of our full spin model with $J_C<0$, we find : $\nu=0.572$ and $\nu=0.577$, see Fig. \ref{fig:limit_antipara}a and c. This agreement confirms our analysis and the relevance of our exact prediction of the streamline geometry.
	\item \textbf{Loop correlation.} We also introduced the correlation $C(r)$, defined as the probability that two edges 
	at a distance $r$ belong to the same loop. 
	This is a crucial observable in loop models \cite{Kondev1995}. For the O($1$) loop model, $C(r) \sim r^{-1/2}$ (the same scaling holds in Ref. \cite{Jaubert2011}).
	Our two sets of simulations shown in Figs. \ref{fig:limit_antipara}b and d confirm that the exact prediction applies to our active hydraulic model. 
	\item \textbf{Loop size distribution.} For critical models, the probability $P_\mathrm{loop}(L)$ that a given loop has size $L$ scales as $P_\mathrm{loop}(L)\sim L^{-\tau}$ \cite{Kondev1995}, with $\tau = 2 + 1/7\simeq 2.12$ for loop O$(1)$ models. 
	This exact result is again in good agreement with our two sets of numerical simulations (\ref{fig:limit_antipara}c and f).
	\item \textbf{Winding loops.} 
	From a more formal perspective, our numerical simulations allow us  to investigate the statistics  of the loops winding around our periodic simulation box.
	We find the average number $n_\mathrm{wind}$ of winding loops is independent of the size of the system: $n_\mathrm{wind} \simeq 0.8$ and that their average length $L_\mathrm{wind}$ scales as $N^{7/4}$ where $N$ is the size of the system (insets of \ref{fig:limit_antipara}a and d). This result agrees with the observations of Ref. \cite{Jaubert2011} and so does the value $d_f = 7/4$ of the fractal dimension.
\end{itemize}
This series of results unambiguously confirm that the geometry of the streamlines in honeycomb active hydraulic networks belongs to the same universality class as the  domain walls of the antiferromagnetic Ising model on the triangular lattice, when deep in the ordered phase and when antiparallel couplings prevail.

\subsubsection{Numerical methods: Monte-Carlo simulation of the  AF Ising model.}
This technical section decribes the numerical method we use to sample the low energy states of the antiferromagnetic Ising model on the triangular lattice (Fig. \ref{fig:example_isingaf}). We use a standard Metropolis Monte-Carlo algorithm \cite{Krauth2006}.
Ising spins $S_f = \pm 1$ are placed on the $N^2$ sites of a periodic triangular lattice of linear size $N$ ($N = 20$ to 1000). The coupling constant $K=\beta J_\mathrm{Ising}$ of the model is set to a value corresponding to a low temperature: $K = 5$.

The initial configuration corresponds to random spins.
At each step, a face $f$ is chosen at random and we compute the energy corresponding to flipping its spin ($S_f \mapsto - S_f$): $\Delta E = -2 K S_f \sum_{f'\sim f} S_{f'}$ where the sum runs over the six neighbors of the spin. 
If $\Delta E<0$, we we change the sign of $S_f$.  $\Delta E \geq 0$, we change the sign of $S_f$ only with probability $e^{-\Delta E}$. Starting from the initial random configuration, we run $1000 N^2$ Monte Carlo steps to thermalize the system. We then run $1000N^2$ additional steps and record the quantities of interest every $N^2$ steps. The results are averaged over 10 independent simulations.

We identify the domain walls as the edges between two sites with opposite spins. The loops are found using a standard
depth-first-search algorithm \cite{Cormen2022}. 
The statistics of the loop lengths and the correlations between edges in the same loop are easily obtained. The gyration radius of a loop in periodic space is computed as follow. 
We consider a loop of $L$ edges centered at positions $(x_1, y_1), \dots, (x_L, y_L)$. We define $x_1' = 0$ and $x_i' = x_{i-1}' + \mathrm{per}(x_i - x_{i-1})$ for $i=2, \dots, L$  where `$\mathrm{per}$' is the shortest distance in periodic space. In other words, we redefine the coordinates iteratively starting from the first site of the loop. 
We proceed similarly for the $y$ coordinates.
The gyration radius $R_g$ of the loop is then defined as
\begin{align}
	R_g &= \sqrt{\mathrm{Var}(x_i') + \mathrm{Var}(y_i')}, &
	\mathrm{Var}(x_i') &= \frac{1}{L} \sum_{i=1}^L x_i'^2 - \left(\frac{1}{L} \sum_{i=1}^L x_i'\right)^2.
\end{align}

\subsection{Geometry of streamlines with no orientational couplings: 3-coloring problem and fully-packed loop O(2) model}  \label{ss:nocoupling}
We now turn to the limit $J_C = 0$ of our active-hydraulics model.
When $J_C = 0$ the streamline loops are self-avoiding but their orientations are uncoupled.
This situation is not directly observed in our experiments where the flow couplings are either parallel or antiparallel.
However, our finite-size simulations indicate that in the crossover between the two regimes, when the fractions of parallel and antiparallel couplings are equal, our $\Phi_e$ spin correctly accounts for the stream line geometry when taking $J_C=0$.

In order to gain a deeper insight into this limit, in the limit of asymptotically large system sizes, we proceed to another analogy with frustrated statistical mechanics.
We map the statistical mechanics of our model defined by Eq.~\eqref{smeq:energyFull} on the 3-coloring problem, and the fully-packed O($2$) model. 
We then take advantage of exact predictions from conformal field theory and compare them with our numerical simulations and experimental measurements.

\subsubsection*{Active hydraulics, 3-coloring model and loop O(2) model: exact results}
\begin{figure}
	\centering
	\includegraphics[width=0.6\textwidth]{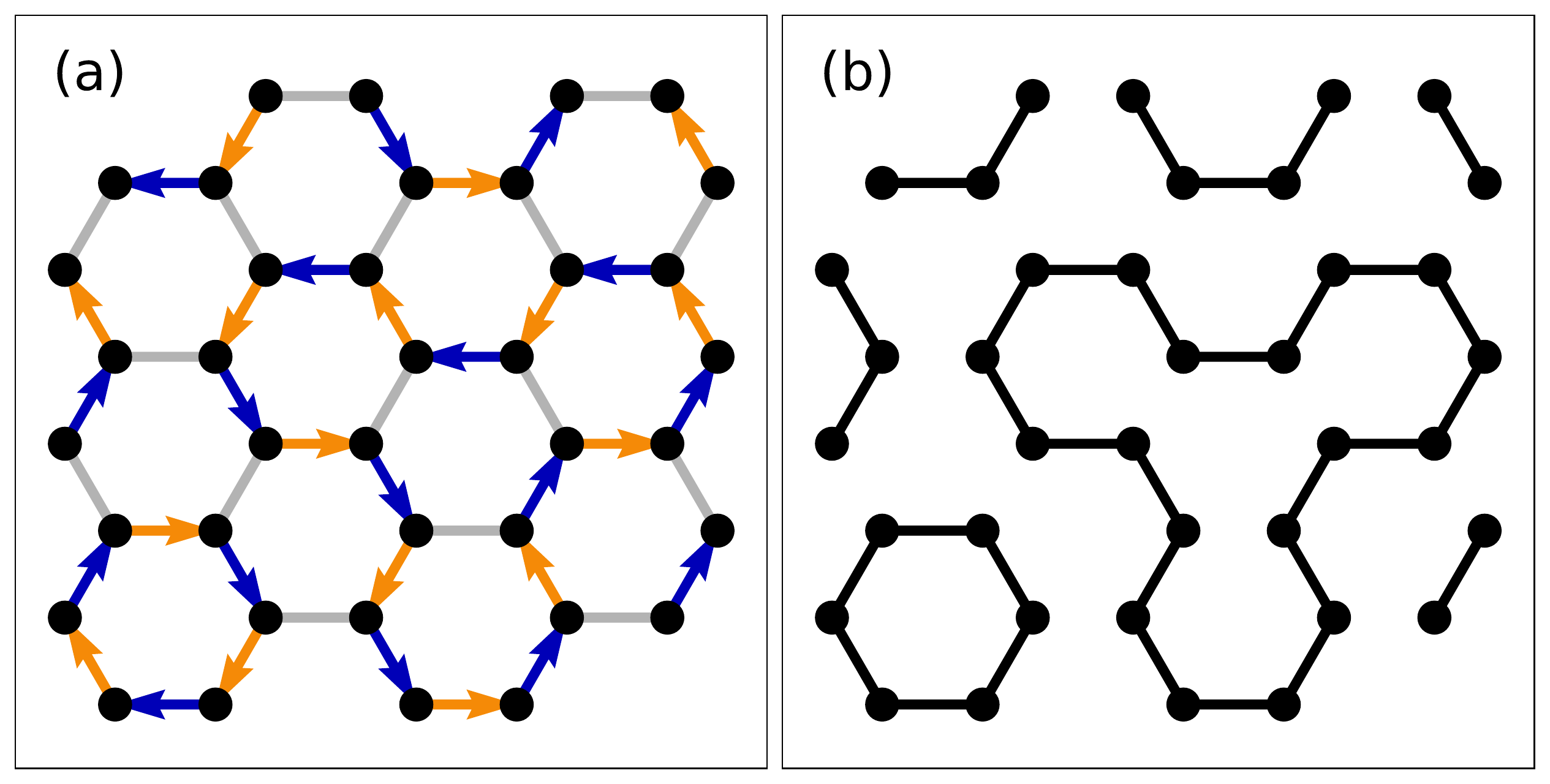} 
	\caption{\textbf{No flow coupling ($J_C=0)$: Mapping on the 3-coloring problem.}
		Under the constraint of maximum flow ($J_A\to \infty$) and no flow coupling ($J_C = 0$), our full spin model (Eq. \eqref{smeq:energyFull}) reduces to the coloring of the edges of a honeycomb lattice with three colors: blue, red and gray, with the three edges linked to a different vertex having different colors. 
		\textbf{(a)} Example of configuration for a periodic system, showing the 3-coloring of the edges.
		\textbf{(b)} This system is equivalent to a fullly packed $O(2)$ loop model (shown for the blue-orange loops).
	}
	\label{fig:example_3colors} \vspace{0.5cm}
	\centering
	\includegraphics[width=\textwidth]{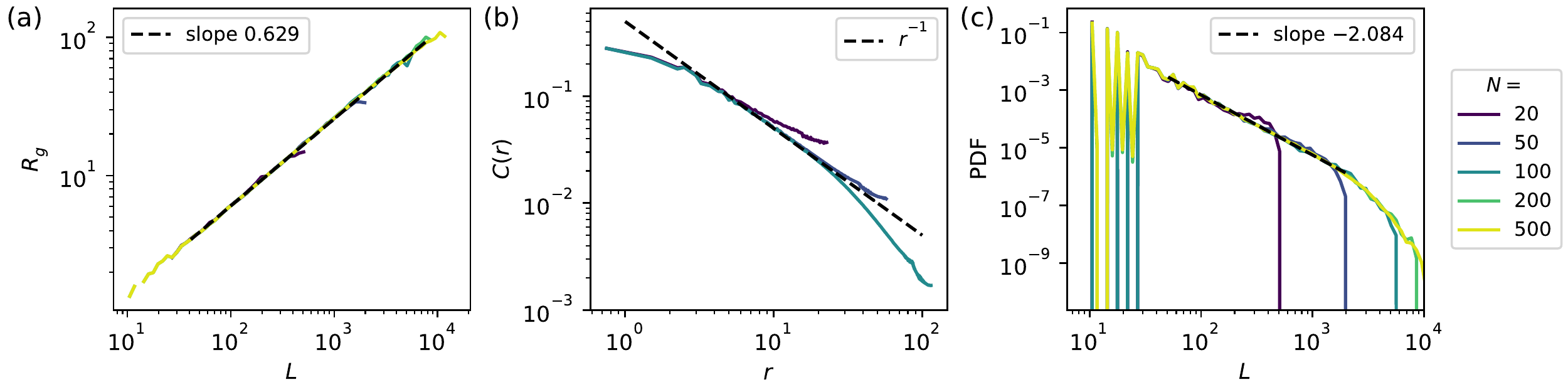}
	\caption{\textbf{No flow coupling ($J_C=0)$: numerical results.}
		\textbf{(a)} Gyration radius of the non-winding loops versus size. We find an exponent $\nu \simeq 0.634$ close to the theoretical value $\nu = 2/3$.
		\textbf{(b)} The decay of the correlation $C(r)$ is consistent with the $r^{-1}$ decay predicted theoretically.
		\textbf{(c)} Probability distribution of the loop sizes. The decay exponent $-2.08$ should be compared with the theoretical prediction $-7/3\approx -2.3$.
	}
	\label{fig:nocoupling_limit}
\end{figure}

We proceed as in the previous section, we focus on the maximally flowing regime: $\beta J_A\to\infty$ in Eq. \eqref{smeq:energyFull}. 
In this limit the vertices of our $\Phi_e$-spin model are coupled to edges with spin configurations of the type $\{+1, -1, 0\}=\{\text{blue}, \text{orange}, \text{gray}\}$.
Since there is no additional constraint constraint ($J_C=0$), our model then exactly corresponds to a 3-coloring model defined on the edges of the hexagonal lattice  the edges, see Fig. \ref{fig:example_3colors}a.
This classical statistical mechanics model was first 
introduced by Baxter in Ref.~\cite{Baxter1970}. 
It was later shown this 3-coloring model is equivalent to the fully packed loop O(2) model~\cite{Kondev1996}.
We recall that $n$ can be interpreted as the fugacity of the loops in a loop model. The value $n=2$ can then be intuitively understood as there are two ways to orient each loop whose conformation are unaffected by their orientation when $J_C=0$.
This situation contrasts with the limit $J_C\ll0$ where 
the orientation of a given loop fully determined the orientation of all the others, thereby leading to a fugacity $n=1$.

The 3-coloring on the honeycomb lattice, and the fully-packed loop O(2) model have been extensively studied in the mathematical physics community, and a number of exact results are available. For instance, for the 3-coloring model, the degeneracy of the number of configurations $\Omega$ is known to scale extensively with the number $N_v$ of vertices: $\ln\Omega = N_v \ln W$ with $W \approx 1.20872$~\cite{Baxter1970}. 
We also recall that the fully-packed loop O(2) model is commonly believed to follow the same statistics as the $\mathrm{SLE}(\kappa=4)$ process \cite{Kager2004}. 
We build on this knowledge to make quantitative predictions on the stream line geometry and test these exact predictions against numerical simulations of our $\Phi_e$-spin model and experiments.

\begin{itemize}
	\item \textbf{Gyration radius.} The fractal dimension of the loops in the fully-packed loop O($2$) model is $d_f = {3}/{2}$ \cite{Kondev1996}: the gyration radius scales as $R_g \sim L^{\nu}$ with $\nu = 1/d_f = 2/3$. 
	This prediction in very good agreement with the numerical simulations of our spin model at $J_C = 0$, see Fig. \ref{fig:nocoupling_limit}a, and is consistent with the value of $\nu$ measured in the crossover regime in our experiments.
	\item \textbf{Loop correlation.} The fully packed loop O(2) model, which has central charge $c=2$ features algebraic correlations. 
	The probability $C(r)$ that two edges at distance $r$ from one another belong to the same loop scales as $C(r) \sim r^{-1}$ \cite{Kondev1995}. 
	This result agrees with our numerical simulations, see Fig. \ref{fig:nocoupling_limit}b.
	\item \textbf{Distribution of loop sizes.} The probability $P_\mathrm{site}(L)$ that the loop passing though a given vertex has size $L$ scales as $P_\mathrm{site}(L) \sim L^{-{(\tau-1)}}$ with $\tau = 7/3$ \cite{Kondev1995,Kondev1996}. The probability $P_\mathrm{loop}(L)$ that a  loop has size $L$ scales as $P_\mathrm{loop}(L) \sim L^{-{\tau}}$.
	These two predictions again agree with our numerical simulations as seen in Fig. \ref{fig:nocoupling_limit}c.
\end{itemize}

In conclusion, our active-hydraulics model belongs to the universality class of  the fully-packed loop O($2$) model in the limit $J_C=0$. 
In particular, the exact value of gyration radius exponent is $\nu=2/3$, which lies in the crossover between the two regimes observed experimentally, Fig. \ref{fig:compa_expe}d.i).
This penultimate set of results further confirm the relevance of our $\Phi_e$-spin model to account for active flows in trivalent networks.

\subsection{Parallel flow couplings: subextensive degeneracy and metastable states} \label{ss:para}
In this last subsection, we focus on the parallel regime. It corresponds to the largest aspect ratios in our experiments and to the limit  $J_C \gg 0$ of our model. 
Unlike the other two cases studied in the above sections ($J_C \ll 0$ and $J_C=0$), we show that the ground state of the model does not map on a critical loop model. 
The ground state configurations have  a sub-extensive degeneracy and cannot be reached in neither the experiments nor Monte-Carlo simulations. 
The configurations that we  observe correspond to local minima of $\mathcal H$,  which we characterize numerically.

\subsubsection{Color-dependent interactions and non-critical phase}
When introducing a non-zero flow coupling $J_C\neq 0$ in   Eq. \eqref{smeq:energyFull}, in the fully packed limit $J_A\to\infty$, our model corresponds to the 3-coloring model with color-dependent interactions.
This class of model was introduced in Ref.~\cite{Verpoort2018}. 
The authors argued that the phase $J_C \gg 0$ is ``frustrated'', ``non-critical'' and ``massive''. 
Instead of 
using this  field theoretic terminology, we instead perform numerical simulations and address the similarities and differences with the other two phases ($J_C<0$ and $J_C=0$).

\begin{figure}
	\centering
	\includegraphics[width=0.7\textwidth]{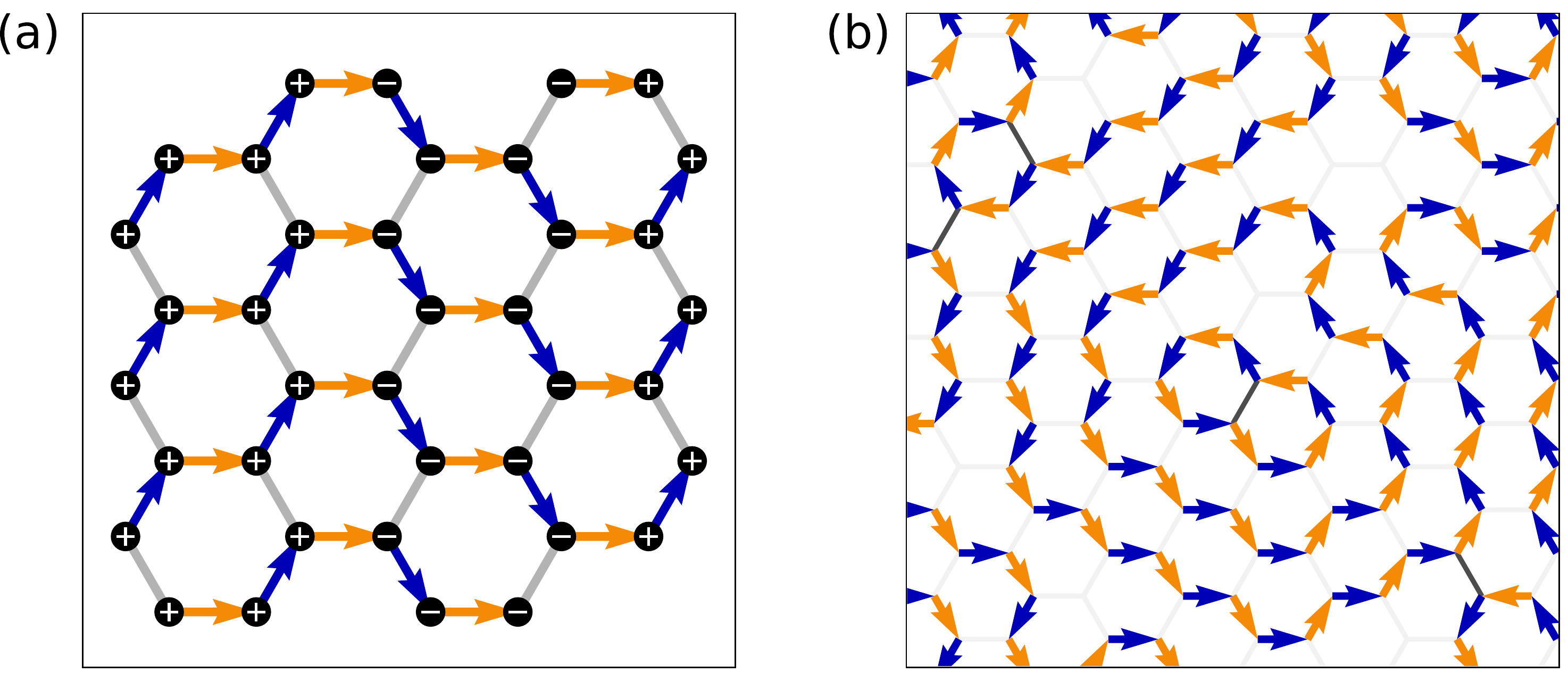} 
	\caption{\textbf{Parallel flow coupling ($J_C > 0$): ground states and final configurations.}
		\textbf{(a)} Example of a ground state configuration of our model when $J_C > 0$. The two constraints are (i) the loops are fully-packed (blue-red-gray at each vertex) and (ii) all couplings are parallel ($\sigma_i\sigma_j = +1$ for each gray edge). They imply that the lowest energy configurations correspond to parallel streamlines that wind around the simulation box.
		\textbf{(b)} A  configuration reached by our numerical simulation. 
		It is not a ground state configuration of $\mathcal H$. It however features a collection of nested loops.
	}
	\label{fig:example_para}
	\vspace{0.5cm}
	\centering
	\includegraphics[width=\textwidth]{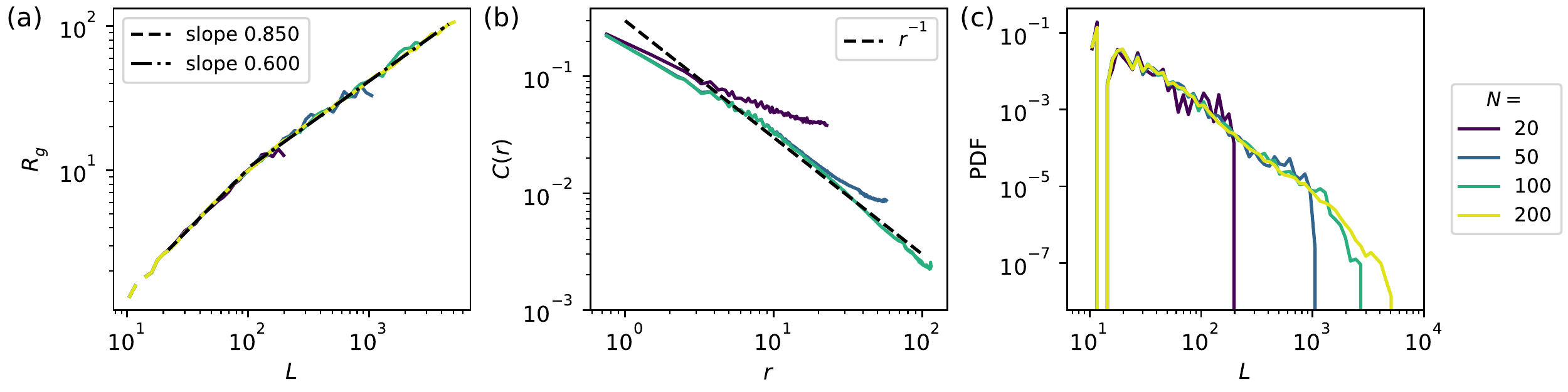}
	\caption{\textbf{Parallel flow coupling ($J_C > 0$): numerical results.}
		\textbf{(a)} Gyration radius versus loop size. We uncover two regimes: for short loops ($L< 100$) the exponent is $\nu_1\approx 0.85$ while for large loops we find $\nu_2\approx 0.60$. 
		\textbf{(b)} The correlation $C(r)$ is consistent with a $r^{-1}$ decay.
		\textbf{(c)} Distribution of loop sizes.
	}
	\label{fig:limit_para}
\end{figure}

We investigate the fully-packed loop configurations ($2/3$ of the edges are flowing) in the limit $J_C\gg0$ where all flow couplings are parallel ($\sigma_i\sigma_j = +1$ accross edges hosting a zero spin $\Phi_e$). 
The only solutions correspond to lines winding around the periodic box, and parallel to one another as illustrated in Fig. \ref{fig:example_para}a. 
In a lattice of $N\times N$ hexagons, the number $\Omega$ of such configurations is of the order of $2^N$ \cite{Verpoort2018}. 
This can be seen by starting with a set of streamlines, or $\Phi_e$ spins pointing along the same direction, say from left to right. To deform the loops,
we can start from the leftmost edge, see Fig.~\ref{fig:example_para} and flip the direction of the streamline, the second edge must host a current flowing to the right hand side, the streamline can then bifurcate upwards or downwards and so forth 
until the streamline winds around the box. This procedure implies that the number of configurations scales as the number of directed random walks which yields a $2^N$ scaling. 
This reasoning shows that the ground state degeneracy (more accurately the ground state complexity) is sub-extensive in the number of vertices $N_v= 3N^2$ (or number of edges) since $\ln\Omega \sim \sqrt{N_v} \ln 2$. 
This behavior contrasts with the $\ln\Omega \sim N_v \ln W$ scaling  found for the 3-coloring model when $J_C=0$, and with the algebraic scaling of the frustrated antiferromagnetic Ising model.

We stress that in a non-periodic system, the lowest energy configurations correspond to fully nested loops, whose degeneracy is also sub-extensive, following the same reasoning.
However,  both in periodic and finite systems these configurations are very hard to reach starting from random initial conditions. 
At low temperature, the worm Monte Carlo algorithm reaches highly stable local energy minima, see Fig.~\ref{fig:example_para}b. 
The same phenomenon happens in the experiments in which we never observed a set of perfectly nested streamlines configurations, see Fig.~\ref{fig:compa_expe}e.i where $f_\mathrm{para} \approx 0.6$ is still much lower than $1$.
A full study of this potentially glassy phase is beyond the objectives of this article, but we  provide  numerical observations which shed some light on the geometry of the  streamlines, i.e of the ground state of the $\Phi_e$-spin model.

\subsubsection{Streamline geometry and non-critical phase for parallel couplings}

We investigate again the same geometrical  quantities numerically. 
\begin{itemize}
	\item \textbf{Gyration radius.} If we were probing ground state configurations, the gyration radius exponent would be trivially given by: $\nu=1$. 
	In fact, the geometry which we observe is more complex. In the metastable states that we probe, Fig. \ref{fig:limit_para}a, we can distinguish two asymptotic behaviors: for short loops ($L<100$), the exponent is $\nu_1\simeq 0.85$ while for long loops ($L>100$) the gyration radius has a scaling $\nu_2 \simeq 0.6$. 
	The impossibility to describe the loops with a single exponent reflect that we are not dealing with a realization of a critical loop model. 
	In the experiments, we probe loops up to size 300. This explains why we find an effective exponent $\nu_\mathrm{exp}\simeq 0.75-0.8$, see Fig. \ref{fig:compa_expe}d.i.
	\item \textbf{Loop correlation.} The probability $C(r)$ that two sites as a distance $r$ belong to the same loop is consistent with a $r^{-1}$ decay (Fig. \ref{fig:limit_para}b). This  scaling is identical to the simulations where $J_C=0$, see Section \ref{ss:nocoupling}. 
	But this scaling is a poor marker of different universality classes. 
	The same scaling would indeed be observed in the ground state where all the streamlines would be parallel.
	Indeed, considering a given edge in a ground state configuration, Fig. \ref{fig:limit_para}a, there are $2$ edges at a distance $r$ that are in the same loop, out of a total number of edges that scales as $r$ ( perimeter of a circle of radius $r$ scales).
	In the ground states we therefore also find $C(r)\sim r^{-1}$. 
	Our numerical simulations tell us that the same scaling holds for the metastable states that we probe.
\end{itemize}

In conclusion, the metastable states of the parallel regime ($J_C>0$) are characterized by a gyration radius exponent $\nu_1\simeq 0.85$ for short loops and a loop correlation $C(r)$ that seems to decay as $r^{-1}$. 
Both observations are consistent with the experiments (Fig. \ref{fig:compa_expe} c.i and d.i).

\section{Finite element simulations}
Our goal is to determine whether the defect fractionalization seen in the flow field  at the subchannel scale is specific to colloidal-roller fluids or generic to polar active matter. 
To answer this question we resort to numerical resolutions of Toner-Tu Hydrodynamics.
Toner-Tu equations, Eqs.~\eqref{exp:masscons} and ~\eqref{exp:TT} are  the equivalent of the Navier-Stokes for polar active fluids, flocks of self-propelled bodies~\cite{Toner1995,Toner1998}. 
In their simplest form they reduce to~\cite{Chardac2021}:
\begin{align}
	&\partial_t \rho + \mathbf{\nabla} \cdot (\rho \mathbf{v}) = 0,
	\label{exp:masscons}\\
	&\partial_t \mathbf{v} + \lambda \mathbf{v} \cdot \mathbf{\nabla} \mathbf{v} = (\alpha -\beta v^2) \mathbf{v} + D \mathbf{\nabla}^2\mathbf{v} - \sigma \mathbf{\nabla}\rho,
	\label{exp:TT}
\end{align}
where $\rho(\mathbf r,t)$ and $\mathbf v(\mathbf r,t)$ are the local density and velocity fields of the polar active liquid. 
We choose the hydrodynamic constants to match the experimental values estimated in Refs \cite{geyer2018,Chardac2021,supekar2023}:  $\lambda = 0.7$, $\sigma = 5 \rm~ mm^2 s^{-2}$, $D \approx 10^{-3}\rm~ mm^2 s^{-1}$, $\alpha_0 = 10^2 \rm~ s^{-1}$, $\rho_c = 3 \times 10^{-3}$, $\beta = 10 \rm~ mm^{-2} s$ (we opt for a definition where $\rho$ is nondimensional as in ~\cite{Chardac2021}). 
The physical meaning of each term has already been thoroughly discussed in the litterature, see e.g.~\cite{Toner1998}.
To solve this set of partial differential equations, we use the  Finite Element Method described in Ref. \cite{Chardac2021}. 
In practice, we use the open source software package FENICS~\cite{logg2012}.

To investigate the structure of the vortical flows in channels of different anisotropy,  we solve the equations in cigar-shaped boxes. The shape consists of a rectangular body with two semicircular segments attached to opposite sides. The rectangular body has a width of $W$ and a length of $L$. The two semicircles have a diameter equal to $W$.
We use  the Babuska penalty method, (see \cite{babuvska1973}), to set tangent boundary conditions for the velocity field: $\mathbf{v}\cdot \mathbf{n} \approx 0$ at each boundary, where $\mathbf{n}$ is the normal vector to the mesh boundary. 
There is no boundary condition for the density field $\rho$, but we checked that mass is globally conserved.
In all simulations, the density is initialized as a constant $\rho_0 = 0.3$, and the velocity vectors are initialized with a random orientation and a norm close to $\sqrt{\alpha / \beta}$.

The time step between two time increments $\delta t$ is chosen to be small compared  to the typical relaxation timescale of the fast-speed mode $ \tau = \alpha(\rho_0) \approx 10^{-1} ~\rm s$. 
In practice, we take $\delta t \approx 10^{-3}\rm ~s$. 
The computational mesh consists of $\sim 2000$ triangular cells. 
In our simulations $L$ varies from $0$  to $200 ~ \rm \mu m$ while $W$ is kept constant and equal to $200 ~ \rm \mu m$.
We interpolate the density and velocity fields using second-order polynomials on Lagrange finite-element cells (see~\cite{logg2012}).  %

\section{Supplementary Videos}
\begin{itemize}
	\item {{\bf Supplementary Video 1} : This video shows the transient dynamics of a colloidal roller fluid flowing through a network of channels. Large scale density and velocity fluctuations relax to reach a homogeneous steady state. Colloid fraction $\sim 30\%$. Channel width: $200\,\mu \rm m$. Channel length: $200\,\mu \rm m$}. Slowed by a factor of 13.
	
	\item {{\bf Supplementary Video 2} : This zoom-out video shows  the  steady-state  flow of a colloidal roller fluid in a honeycomb network of channels. Colloid fraction $\sim 30\%$. Channel width: $200\,\mu \rm m$. Channel length: $220\,\mu \rm m$}. Slowed by a factor of 6.5.
	
	\item {{\bf Supplementary Video 3} : This large scale video shows the  steady-state  flow of a colloidal roller fluid in a honeycomb network. Laminar flows are frustrated at every node of the network. Colloid fraction $\sim 30\%$. Channel width: $200\,\mu \rm m$. Channel length: $120\,\mu \rm m$}. Slowed by a factor of 13.
	
	\item {{\bf Supplementary Video 4} : This video shows a zoom from the  steady-state  flow of a colloidal roller fluid at a trivalent node. The node has a flow splitter. The flows are laminar in two channels linked to the left node. The third channel host two counter rotating vortices. Colloid fraction $\sim 30\%$. Channel width: $200\,\mu \rm m$. Channel length: $200\,\mu \rm m$}. Slowed by a factor of 13.
	
	\item {{\bf Supplementary Video 5} : This video shows a zoom from the  steady-state  flow of a colloidal roller fluid at a trivalent node. The node has a flow splitter. The three channels linked to the left node host vortices. Colloid fraction $\sim 30\%$. Channel width: $200\,\mu \rm m$. Channel length: $200\,\mu \rm m$}. Slowed by a factor of 13.
	
	\item {{\bf Supplementary Video 6} : This video shows a zoom from the  steady-state  flow of a colloidal roller fluid at a trivalent node. The node is a simple trivalent junction without splitter. The flows are laminar in two channels linked to the left node. The third channel host two counter rotating vortices. Colloid fraction $\sim 30\%$. Channel width: $200\,\mu \rm m$. Channel length: $220\,\mu \rm m$}. Slowed by a factor of 13.

\end{itemize}

\section{Supplementary Figure}

\begin{figure}[h!]
	\centering
	\includegraphics[width=0.5\textwidth]{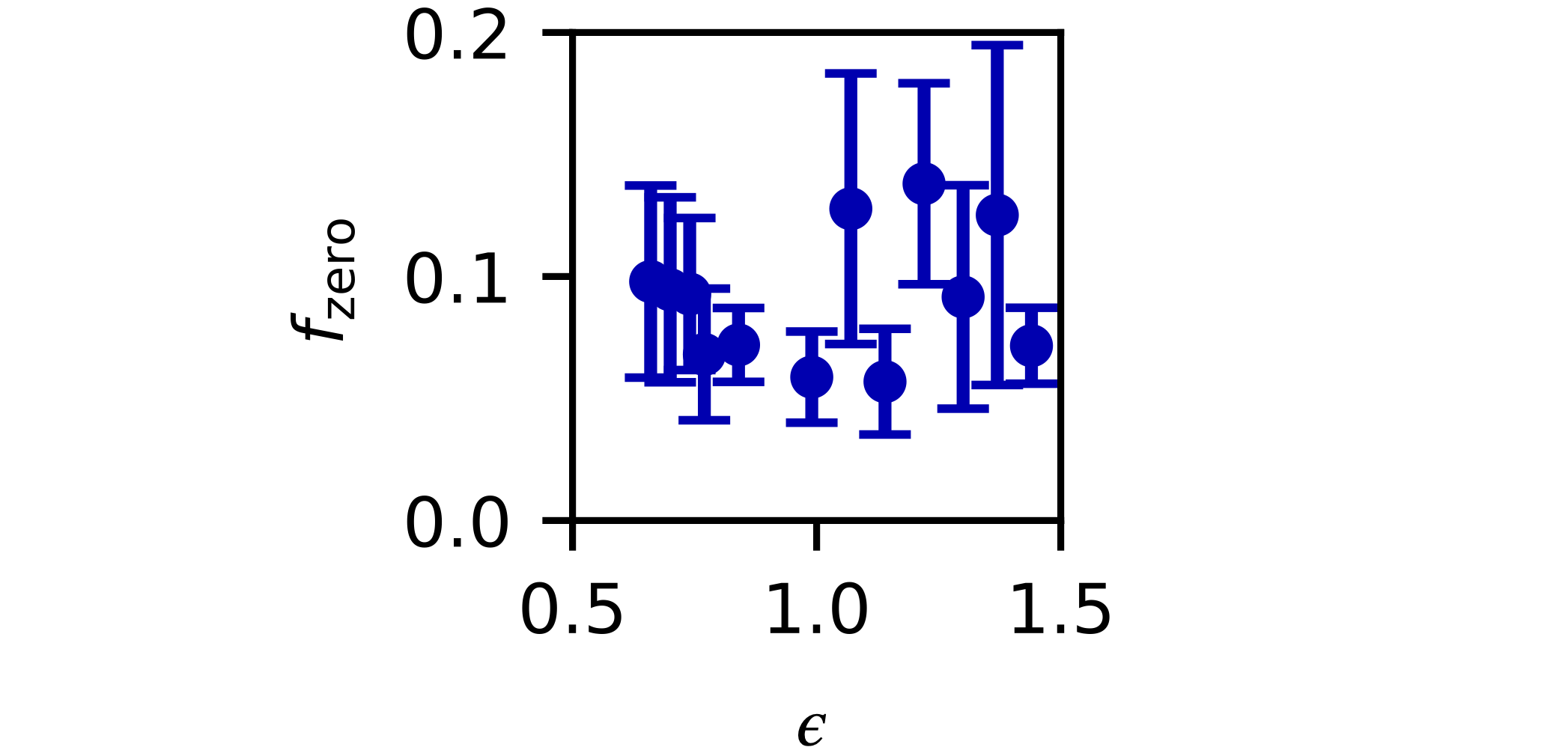}
	\caption{{\bf Fraction of nodes with three vanishing fluxes.} The fraction $f_{\rm zero}$ does not change significantly with $\epsilon$ and remains around $0.1$. The error bars correspond to the $95\%$ confidence interval.}
	
\end{figure}

\end{document}